\documentclass[12pt]{article}
\usepackage{graphicx}
\usepackage{amsmath}
\usepackage{amsfonts}
\usepackage{amssymb}

\newtheorem{proposition}{Proposition}
\newtheorem{property}{Property}
\newenvironment{proof}[1][Proof]{\textbf{#1.} }{\ \rule{0.5em}{0.5em}}
\begin{document}

\title{\textbf{An exact fluid model for relativistic electron beams: The many moments case.}}
\author{M.C. Carrisi, S. Pennisi}
\date{}
\maketitle
 \small {\em \noindent Dipartimento di Matematica ed Informatica,
Universit\`{a} di Cagliari, Via Ospedale 72,\,\ 09124 Cagliari, Italy; e-mail:
cristina.carrisi@tiscali.it;spennisi@unica.it}
 \\
~\\
An interesting and satisfactory fluid model has been proposed in literature for the the
description of relativistic electron beams. It was obtained with 14 independent variables
by imposing the entropy principle and the relativity principle. Here the case is
considered with an arbitrary number of independent variables, still satisfying the above
mentioned two principles; these lead to conditions whose general solution is here found.
We think that the results satisfy also a certain ordering with respect to a smallness
parameter $\epsilon$ measuring the dispersion of the velocity about the mean; this
ordering generalizes that appearing in literature for the 14 moments case.
\section{Introduction}
Amendt and Weitzner proposed in \cite{1} and \cite{2} the following system of
quasi-linear partial differential equations
\begin{equation}\label{p1}
\partial_{\alpha}V^{\alpha}=0,\qquad
\partial_{\alpha}T^{\alpha\beta}=eF^{\beta\alpha}V_{\alpha},\qquad
\partial_{\alpha}A^{\alpha\beta\gamma}=2eT_{\alpha}^{(\beta}F^{\gamma)\alpha},
\end{equation}
to describe the behavior of relativistic electron beams, where
$V^{\alpha}$ is the particle flux vector, $T^{\alpha\beta}$ is the
energy-momentum tensor , $F^{\beta\alpha}$ is the electromagnetic
field tensor, $e$ is the electron charge and
$A^{\alpha\beta\gamma}$ represents the tensor of fluxes. Let's
consider the counterpart of the above variables in statistical
mechanics. They are defined as moments of the distribution
function $f(x^{\alpha},p^{\alpha})$,
\begin{equation}\label{p2}
V^{\alpha}=\int f p^{\alpha}dP, \qquad T^{\alpha\beta}=\int f p^{\alpha}p^{\beta}dP,
\qquad A^{\alpha\beta\gamma}=\int f p^{\alpha}p^{\beta}p^{\gamma}dP,
\end{equation}
where $p^{\alpha}$ is the four-momentum of the particle so that we have
$p^{\alpha}p_{\alpha}=-m^2$ and $dP=\sqrt{-g}\frac{dp_1dp_2dp_3}{p_0}$ is the invariant
element of the momentum space; m is the particle mass. From eqs. $(\ref{p2})$ the
following ``trace condition" holds:
\begin{equation}\label{p3}
A^{\alpha\beta\gamma}g_{\beta\gamma}=-m^2V^{\alpha}.
\end{equation}
The closure proposed by Amendt and Weitzner in \cite{1} and
\cite{2} is covariant, complete and satisfies the above mentioned
ordering with respect to $\epsilon$; the resulting system is
hyperbolic, but it satisfies the trace condition only
approximately. In \cite{3} and \cite{4} a new closure has been
found. It is very satisfactory because it satisfies exactly the
condition $(\ref{p3})$ and leads to a symmetric hyperbolic system,
but imposes the entropy principle only up to second order with
respect to thermodynamical equilibrium. In \cite{5} the same
result has been achieved but up to whatever order. Here the exact
general solution for the \textbf{many moments} case is found,
satisfying all these conditions up to whatever order. Extension to
very many moments are needed in order to improve on the results of
ordinary thermodynamics, as shown in \cite{6}, page 197. Moreover,
we will exploit general equations so that the results may be
applied also
to other materials even if different from electron beams.\\
For the case with many moments we have to choose an even number M and an odd number N;
after that  the equations are
\begin{equation}\label{1}
  \begin{cases}
   \partial_{\alpha}A^{\alpha\alpha_{1}...\alpha_{M}}=I^{\alpha_{1}...\alpha_{M}}, \\
   \partial_{\alpha}B^{\alpha\alpha_{1}...\alpha_{N}}=I^{\alpha_{1}...\alpha_{N}}.
  \end{cases}
\end{equation}
Equations involving lower order tensors are already included in $(\ref{1})$ because of
the following trace conditions $(\ref{4a1})$.\\
All the tensors appearing in the above equations are symmetric and M+N is odd in order to
obtain independent equations. The counterparts of these variables in statistical
mechanics are
\begin{equation*}
A^{\alpha\alpha_{1}...\alpha_{M}}=\int f
p^{\alpha}p^{\alpha_1}\cdots p^{\alpha_M}dP, \qquad
B^{\alpha\alpha_{1}...\alpha_{N}}=\int f
p^{\alpha}p^{\alpha_1}\cdots p^{\alpha_N}dP,
\end{equation*}
from which the trace conditions follows
\begin{eqnarray}\label{4a1}
-m^2
A^{\alpha\alpha_{1}...\alpha_{M-2}}=A^{\alpha\alpha_{1}...\alpha_{M}}g_{\alpha_{M-1}\alpha_M},\qquad
-m^2
B^{\alpha\alpha_{1}...\alpha_{N-2}}=B^{\alpha\alpha_{1}...\alpha_{N}}g_{\alpha_{N-1}\alpha_N}.
\end{eqnarray}
Let us define the maximum trace of a tensor as the trace of the trace ... of the trace of
this tensor, so many times as possible.  The maximum traces of
$I^{\alpha_{1}...\alpha_{M}}$ and of $I^{\alpha_{1}...\alpha_{N}}$ are zero, so that the
maximum traces of eqs. $(\ref{1})$ are the conservation laws of
mass and of momentum-energy. \\
Now there are less independent components in the eqs. $(\ref{1})$
than in the variables $ A^{\alpha\alpha_{1}...\alpha_{M}}$ and
$B^{\alpha\alpha_{1}...\alpha_{N}}$, so that relations between
these variables are needed; aim to find them is called ``closure
problem". We investigate this problem following a commonly used
procedure in Extended Thermodynamics, i.e., to impose the
supplementary conservation law
\begin{equation}\label{4a}
\partial_{\alpha}h^{\alpha}=\sigma\geq 0,
\end{equation}
that must hold true for all the solutions of the system
$(\ref{1})$. As a bonus we will obtain a symmetric hyperbolic
system with nice mathematical properties, such as well-posedness
of the Cauchy problem and continuous dependence on the initial
data (see \cite{7}). By using a property proved by Friedrichs, Lax
and Liu \cite{8}, \cite{9}, \cite{10}, we obtain that eq.
$(\ref{4a})$ amounts in assuming the existence of parameters
$\lambda_{\alpha_1\cdots\alpha_M}$ e
$\mu_{\alpha_1\cdots\alpha_N}$, called Lagrange Multipliers, such
that
\begin{equation}\label{2'}
dh^{\alpha}=\lambda_{\alpha_1\cdots\alpha_M}dA^{\alpha\alpha_{1}...\alpha_{M}}+
\mu_{\alpha_1\cdots\alpha_N}dB^{\alpha\alpha_{1}...\alpha_{N}},
\end{equation}
\begin{equation*}
\lambda_{\alpha_1\cdots\alpha_M}I^{\alpha_{1}...\alpha_{M}}+
\mu_{\alpha_1\cdots\alpha_N}I^{\alpha_{1}...\alpha_{N}}\geq 0,
\end{equation*}
where $h^{\alpha}$ is the entropy-(entropy flux) 4-vector.\\
We follow now an idea developed by Boillat, Ruggeri and Strumia
which they applied in different physical contexts concerning both
the classical \cite{11} and the relativistic case \cite{12},
\cite{13}. In other words, we define
\begin{equation}\label{2''}
h^{'\alpha}=-h^{\alpha}+\lambda_{\alpha_1\cdots\alpha_M}A^{\alpha\alpha_{1}...\alpha_{M}}+
\mu_{\alpha_1\cdots\alpha_N}B^{\alpha\alpha_{1}...\alpha_{N}},
\end{equation}
and take the Lagrange Multipliers as independent variables. In this way eq.
$(\ref{2'})_1$ becomes
\begin{equation*}
dh^{'\alpha}=A^{\alpha\alpha_{1}...\alpha_{M}}d\lambda_{\alpha_1\cdots\alpha_M}+
B^{\alpha\alpha_{1}...\alpha_{N}}d\mu_{\alpha_1\cdots\alpha_N},
\end{equation*}
from which
\begin{equation}\label{2}
A^{\alpha\alpha_{1}...\alpha_{M}}=\frac{\partial
h^{'\alpha}}{\partial \lambda_{\alpha_1\cdots\alpha_M}},\qquad
B^{\alpha\alpha_{1}...\alpha_{N}}=\frac{\partial
h^{'\alpha}}{\partial \mu_{\alpha_1\cdots\alpha_N}}.
\end{equation}
In this way the tensors appearing in the balance equations
$(\ref{1})$ are found as functions of the pa\-ra\-me\-ters
$\lambda_{\alpha_{1}...\alpha_{M}}$ and
$\mu_{\alpha_{1}...\alpha_{N}}$, called also mean field, as soon
as $h^{'\alpha}$ is known. So it remains to find the exact general
expression for $h^{'\alpha}$ such that both members of eq.
$(\ref{2})$ are symmetric; some steps in this direction will be
done in section 3. It will be necessary to distinguish between the
cases N=1 and $N>1$; the first of these will be analyzed in
section 5, the other in section 6. Their treatment needs some
useful theorems which will be proved in section 4. In the next
section we will consider the definition of equilibrium and the
implication of the above conditions in this state. At last
conclusions will be drown.\\
We consider necessary to mention also the important context of
which the present paper becomes a part, i.e., Extended
Thermodynamics. To this end we conclude this section by citing the
pioneering papers \cite{14} for the classical context and
\cite{15} for the relativistic one, and also the important book
\cite{6} which contains many results in this context. We reccomand
to reed also the important paper \cite{16} by Geroch and
Lindbloom.
\section{Definition of equilibrium and properties}
Equilibrium is defined as the state described by the independent variables $\lambda$ and
$\mu_{\alpha}$ such that
\begin{eqnarray}\label{8a}
 \lambda_{\beta_1\cdots \beta_M}=\lambda g_{(\beta_1\beta_2}\cdots g_{\beta_{M-1}\beta_M)}(-m^2)^{-\frac{M}{2}},\nonumber \\
 \mu_{\beta_1\cdots \beta_N}=\mu_{(\beta_1} g_{\beta_2\beta_3}\cdots
 g_{\beta_{N-1}\beta_N)}(-m^2)^{-\frac{N-1}{2}},
\end{eqnarray}
from which it follows
\begin{eqnarray}\label{8b}
\lambda=2\frac{(M-1)!!}{(M+2)!!} \lambda_{\alpha_1\cdots \alpha_M} g^{\alpha_1\alpha_2}\cdots g^{\alpha_{M-1}\alpha_M}(-m^2)^{\frac{M}{2}},\nonumber \\
\mu_{\alpha}=8\frac{N!!}{(N+3)!!}\mu_{\alpha\alpha_1\cdots
\alpha_{N-1}}g^{\alpha_1\alpha_2}\cdots
 g^{\alpha_{N-2}\alpha_{N-1}}(-m^2)^{\frac{N-1}{2}}.
\end{eqnarray}
The physical meaning of this definition is evident when we substitute eqs. $(\ref{8a})$
into eq. $(\ref{2'})_1$; by using also the trace conditions $(\ref{4a1})$, eq.
$(\ref{2'})_1$ becomes
\begin{equation*}
dh^{\alpha}=\lambda dA^{\alpha}+\mu_{\alpha_1}dB^{\alpha\alpha_1}
\end{equation*}
which is still eq. $(\ref{2'})_1$, but in the case M=0, N=1, i.e. in the case we consider
only the conservation laws of mass and of momentum-energy.\\
It is also evident that eqs. $(\ref{8b})$ are identities if M=0, N=1. (Use firstly the
identity $(M-1)!!=\frac{(M+1)!!}{M+1}$).\\
Eqs. $(\ref{2})$ now become
\begin{equation*}
A^{\alpha}=\frac{\partial h^{'\alpha}}{\partial \lambda}, \qquad
B^{\alpha\alpha_1}=\frac{\partial h^{'\alpha}}{\partial
\mu_{\alpha_1}}
\end{equation*}
and the symmetry condition on the second of these is surely
satisfied; in fact, from the representation theorems \cite{7},
\cite{8} we have that
\begin{equation}\label{z1}
h^{'\alpha}=H(\lambda,\gamma)\mu^{\alpha}
\end{equation}
with $\gamma=\sqrt{-\mu^{\alpha}\mu_{\alpha}}$ and $H(\lambda,\gamma)$ is an arbitrary
function. It follows
\begin{equation}\label{z3}
A^{\alpha}=\frac{\partial H}{\partial\lambda}\mu^{\alpha},
\end{equation}
\begin{equation}\label{z2}
B^{\alpha\alpha_1}=-\frac{1}{\gamma}\frac{\partial
H}{\partial\gamma}\mu^{\alpha}\mu^{\alpha_1}+Hg^{\alpha\alpha_1}
\end{equation}
which is surely symmetric.\\
We stress now that the function $H(\lambda,\gamma)$ has to be arbitrary, if we want that
our equations may be applied to all materials. In fact, eqs. $(\ref{z1})$-$(\ref{z2})$
and eq. $(\ref{2''})$ yield
\begin{equation}\label{z4}
h^{\alpha}=-nsu^{\alpha},\quad A^{\alpha}=nu^{\alpha},\quad
B^{\alpha\beta}=eu^{\alpha}u^{\beta}+ph^{\alpha\beta}
\end{equation}
with
\begin{eqnarray}\label{z5}
&& u^{\alpha}=\frac{\mu^{\alpha}}{\gamma},\qquad
h^{\alpha\beta}=g^{\alpha\beta}+u^{\alpha}u^{\beta}\quad
\text{(projector)},\nonumber \\
&& n=\gamma\frac{\partial H}{\partial \lambda}\quad
\text{(particle
density)}, p=H \quad \text{(pressure)},\nonumber \\
&& e=-H-\gamma\frac{\partial H}{\partial\gamma}\quad \text{(energy
density)},\quad s=-\lambda-\gamma\frac{\frac{\partial
H}{\partial\gamma}}{\frac{\partial H}{\partial\lambda}}\quad
\text{(entropy density)}.
\end{eqnarray}
From $(\ref{z5})$ it follows
$d\left(\frac{e}{n}\right)+pd\left(\frac{1}{n}\right)=\frac{1}{\gamma}ds$ which, compared
with the Gibbs relation
\begin{equation}\label{z6}
Tds=d\left(\frac{e}{n}\right)+pd\left(\frac{1}{n}\right)
\end{equation}
identifies $\gamma$ as $\frac{1}{T}$, with $T$ the absolute temperature. Now
$(\ref{z5})_4$ can be used to change variables from $\gamma$, $\lambda$ to $\gamma$, p;
by substituting its solution $\lambda=\lambda(\gamma,p)$ into eqs. $(\ref{z5})_{3,5}$ we
obtain the state functions $n=n(\gamma,p)$ and $e=e(\gamma,p)$. Vice versa, if these
state functions are assigned, by substituting p from $(\ref{z5})_4$ into
$(\ref{z5})_{3,5}$, these become
\begin{eqnarray}\label{z7}
\frac{\partial H}{\partial\lambda}=\frac{1}{\gamma}n[\gamma,H(\lambda,\gamma)],\nonumber \\
\frac{\partial
H}{\partial\lambda}=-\frac{1}{\gamma}e[\gamma,H(\lambda,\gamma)]-\frac{1}{\gamma}H(\lambda,\gamma)
\end{eqnarray}
which are differential equations for the determination of the function
$H(\lambda,\gamma)$.\\
Therefore, if we want that our field equations may be used for all materials, i.e., for
all possible state functions $n=n(\gamma,p)$ and $e=e(\gamma,p)$, then the function
$H(\lambda,\gamma)$ must be arbitrary.\\
Note also that  the integrability condition for eqs. $(\ref{z7})$ is $(e+p)n_p-\gamma
n_{\gamma}=ne_p$. But this is not a new condition on the state function because it is the
same integrability condition for the equations
\begin{eqnarray*}
\frac{\partial s}{\partial p}=\gamma \frac{\partial}{\partial p}\left(\frac{e}{n}\right)+\gamma p \frac{\partial}{\partial p}\left(\frac{1}{n}\right) \\
\frac{\partial s}{\partial \gamma}=\gamma \frac{\partial}{\partial
\gamma}\left(\frac{e}{n}\right)+\gamma p \frac{\partial}{\partial
\gamma}\left(\frac{1}{n}\right)
\end{eqnarray*}
which are equivalent to the Gibbs relation reported above in eq. $(\ref{z6})$.\\
We conclude this section showing that an expression for $h^{'\alpha}$ at equilibrium is
\begin{equation}\label{z8}
h^{'\alpha}=\int F(\lambda,\mu_{\nu}p^{\nu})p^{\alpha}dp
\end{equation}
where the function F(X,Y) is related to the distribution function at equilibrium by
\begin{equation}\label{z9}
\frac{\partial}{\partial X}F(X,Y)=f_{eq}(X,Y).
\end{equation}
The expression $(\ref{z8})$ will be useful in the sequel and is equivalent to
$(\ref{z1})$ with
\begin{equation}\label{z10}
H(\lambda,\gamma)=-\frac{4\pi}{\gamma}m^3\int_0^{\infty}F(\lambda,\gamma m
\cosh\rho)\sinh^2 \rho ~d\rho.
\end{equation}
which gives the relation between H and F.\\
It follows that F(X,Y) is arbitrary, because $H(\lambda,\gamma)$ is arbitrary, and
$(\ref{z8})$ is the most general expression for $h^{'\alpha}$ at equilibrium. A
particular case follows from $(\ref{z9})$ and from
\begin{equation*}
f_{eq}(X,Y)=\frac{\frac{w}{h^3}}{e^{\frac{X+Y}{k}}\pm 1}
\end{equation*}
which is the J\"{u}ttner \cite{19}, \cite{20} distribution
function at equilibrium, where k is the Boltzmann constant, the
upper and lower signs refer to Fermions and Bosons, respectively,
h is the Plank's constant and w is equal to 2s+1 for particles
with spin $\frac{sh}{2\pi}$.
\section{The entropy principle}
To impose eqs. $(\ref{2})$ we have to find the most general expression of $h^{'\alpha}$
such that the left hand sides are symmetric. We will refer to this as ``the symmetry
condition". An exact particular solution of this condition is
\begin{equation*}
h^{'\alpha}_1=\int F(\lambda_{\alpha_{1}...\alpha_{M}}p^{\alpha_{1}}...p^{\alpha_{M}},
\mu_{\beta_{1}...\beta_{N}}p^{\beta_{1}}...p^{\beta_{N}})p^{\alpha}dP,
\end{equation*}
as it can be easily verified, where the function F(X,Y) has been
determined in the previous section (eqs. $(\ref{z7})-(\ref{z10})$)
in terms of the state functions at equilibrium. This solution is
more general than the corresponding one in the kinetic approach
\cite{21}, where the particular case $F(X,Y)=F(X+Y)$ is
considered, but it is
not still the most general one. We aim here to find this most general solution. \\
To this end let us note that the symmetry condition for eqs.
$(\ref{2})$ reads
\begin{equation*}
\frac{\partial h^{'[\alpha}}{\partial\lambda_{\alpha_1]\alpha_2\cdots \alpha_M}}=0,
\qquad \qquad \qquad \frac{\partial h^{'[\alpha}}{\partial\mu_{\alpha_1]\alpha_2\cdots
\alpha_N}}=0;
\end{equation*}
if we subtract from these their expressions with $h^{'\alpha}_1$ instead of
$h^{'\alpha}$, we find
\begin{equation}\label{8c}
\frac{\partial \Delta h^{'[\alpha}}{\partial\lambda_{\alpha_1]\alpha_2\cdots
\alpha_M}}=0, \qquad \qquad \qquad \frac{\partial \Delta
h^{'[\alpha}}{\partial\mu_{\alpha_1]\alpha_2\cdots \alpha_N}}=0.
\end{equation}
with $\Delta h^{'\alpha}=h^{'\alpha}-h^{'\alpha}_1$.\\
But in the previous section we have found that $(h^{'\alpha}_1)_{eq}$ is the most general
expression for $(h^{'\alpha})_{eq}$, so that we have
\begin{equation}\label{8d}
(\Delta h^{'\alpha})_{eq}=0.
\end{equation}
In this way, we have now to find the most general solution of eqs. $(\ref{8c})$ and
$(\ref{8d})$, after that we will have
\begin{equation}\label{8e}
h^{'\alpha}=h^{'\alpha}_1+\Delta h^{'\alpha}.
\end{equation}
To exploit eqs. $(\ref{8c})$ and $(\ref{8d})$, let's firstly consider the Taylor
expansion of $\Delta h^{'\alpha}$ around equilibrium:
\begin{equation}\label{4}
\Delta
h^{'\alpha}=\sum_{h,k=0}^{\infty}\frac{1}{h!k!}C_{h,k}^{\alpha
A_1\cdots A_h B_1\cdots
B_k}\widetilde{\lambda}_{A_1}\cdots\widetilde{\lambda}_{A_h}\widetilde{\mu}_{B_1}\cdots
\widetilde{\mu}_{B_k},
\end{equation}
where the multiindex notation $A_i=\alpha_{i_1}\cdots \alpha_{i_M}$,
$B_j=\alpha_{j_1}\cdots \alpha_{j_N}$ has been used, and, moreover
\begin{equation}\label{5}
C_{h,k}^{\alpha A_1\cdots A_h B_1\cdots
B_k}=\left(\frac{\partial^{h+k}\Delta
h^{'\alpha}}{\partial\lambda_{A_1}\cdots\partial\lambda_{A_h}\partial\mu_{B_1}\cdots
\partial\mu_{B_k}}\right)_{eq.}
\end{equation}
and
\begin{equation}\label{6}
  \begin{cases}
\widetilde{\lambda}_{\beta_1\cdots\beta_M}=\lambda_{\beta_1\cdots\beta_M}-\lambda~ g_{(\beta_1\beta_2}\cdots g_{\beta_{M-1}\beta_M)}(-m^2)^{-\frac{M}{2}}, \\
\widetilde{\mu}_{\beta_1\cdots\beta_N}=\mu_{\beta_1\cdots\beta_N}-\mu_{(\beta_1}
g_{\beta_2\beta_3}\cdots g_{\beta_{N-1}\beta_N)}(-m^2)^{-\frac{N-1}{2}},
  \end{cases}
\end{equation}
denote the deviation of the Lagrange multipliers from their value $(\ref{8a})$ at
equilibrium. Note also that their traces are zero, as consequence of eq. $(\ref{8b})$.\\
Because of the symmetry shown by eqs. $(\ref{8c})$ it is possible
to exchange the index $\alpha$ with each other index taken from
those included in $A_i$ or $B_j$ and moreover each $A_i$ can be
exchanged with each other $A_s$ or $B_r$. So the tensor
$A_{h,k}^{\cdots}$ is symmetric with respect to every couple of indexes.\\
Let's consider $\frac{\partial^{h+k}\Delta
h^{'\alpha}}{\partial\lambda_{A_1}\cdots\partial\lambda_{A_h}\partial\mu_{B_1}\cdots
\partial\mu_{B_k}}$ depending on $\lambda_{\gamma_1\cdots \gamma_M}$ as a composite
function through $\widetilde{\lambda}_{\gamma_1\cdots \gamma_M}$
and $\lambda$; after that let us take its derivative with respect
to $\lambda_{\gamma_1\cdots \gamma_M}$ and calculate the result at
equilibrium. We obtain
\begin{eqnarray}\label{11a}
&&C_{h+1,k}^{\alpha A_1\cdots A_{h+1} B_1\cdots
B_k}=\frac{\partial C_{h,k}^{\alpha A_1\cdots A_h B_1\cdots
B_k}}{\partial \widetilde{\lambda}_{\beta_1\cdots
\beta_M}}\frac{\partial \widetilde{\lambda}_{\beta_1\cdots
\beta_M}}{\lambda_{\gamma_1\cdots \gamma_M}}+\frac{\partial
C_{h,k}^{\alpha A_1\cdots A_h B_1\cdots
B_k}}{\partial \lambda}\frac{\partial \lambda}{\lambda_{\gamma_1\cdots \gamma_M}}= \nonumber\\
 &=& \frac{\partial C_{h,k}^{\alpha
A_1\cdots A_h B_1\cdots B_k}} {\partial \widetilde{\lambda}_{\beta_1\cdots
\beta_M}}\left(g^{(\gamma_1}_{\beta_1}\cdots g^{\gamma_M)}_{\beta_M}- g^{(\gamma_1\gamma_2}\cdots g^{\gamma_{M-1}\gamma_M)}g_{\beta_1\beta_2}\cdots g_{\beta_{M-1}\beta_M}2\frac{(M-1)!!}{(M+2)!!}\right)+\nonumber\\
 &+& \frac{\partial C_{h,k}^{\alpha
A_1\cdots A_h B_1\cdots B_k}} {\partial \lambda} g^{(\gamma_1\gamma_2}\cdots
g^{\gamma_{M-1}\gamma_M)}2\frac{(M-1)!!}{(M+1)!!}(-m^2)^{\frac{M}{2}},
\end{eqnarray}
where $(\ref{8b})_1$ and $(\ref{6})_1$ have been used.\\
Multiplying both sides by $g_{\gamma_1\gamma_2}\cdots g_{\gamma_{M-1}\gamma_M}$ we obtain
\begin{equation}\label{8}
C_{h+1,k}^{\alpha A_1\cdots A_{h}\gamma_1\cdots \gamma_M B_1\cdots
B_k}g_{\gamma_1\gamma_2}\cdots
g_{\gamma_{M-1}\gamma_M}=\frac{\partial C_{h,k}^{\alpha A_1\cdots
A_h B_1\cdots B_k}} {\partial \lambda} (-m^2)^{\frac{M}{2}}.
\end{equation}
Similarly, let's consider $\frac{\partial^{h+k}\Delta
h^{'\alpha}}{\partial\lambda_{A_1}\cdots\partial\lambda_{A_h}\partial\mu_{B_1}\cdots
\partial\mu_{B_k}}$ depending on $\mu_{\gamma\gamma_1\cdots \gamma_{N-1}}$ as a composite
function trough $\widetilde{\mu}_{\gamma\gamma_1\cdots \gamma_{N-1}}$ and $\mu_{\beta}$;
after that let us take its derivative with respect to $\mu_{\gamma\gamma_1\cdots
\gamma_{N-1}}$ and calculate the result at equilibrium. We obtain
\begin{eqnarray*}
&&C_{h,k+1}^{\alpha A_1\cdots A_{h} B_1\cdots
B_{k+1}}=\frac{\partial C_{h,k}^{\alpha A_1\cdots A_h B_1\cdots
B_k}}{\partial \widetilde{\mu}_{\beta\beta_1\cdots
\beta_{N-1}}}\frac{\partial \widetilde{\mu}_{\beta\beta_1\cdots
\beta_{N-1}}}{\mu_{\gamma\gamma_1\cdots
\gamma_{N-1}}}+\frac{\partial C_{h,k}^{\alpha A_1\cdots A_h
B_1\cdots
B_k}}{\partial \mu_{\beta}}\frac{\partial \mu_{\beta}}{\mu_{\gamma\gamma_1\cdots \gamma_{N-1}}}= \\
 &=& \frac{\partial C_{h,k}^{\alpha
A_1\cdots A_h B_1\cdots B_k}} {\partial \widetilde{\mu}_{\beta\beta_1\cdots
\beta_{N-1}}}\left(g_{\beta}^{(\gamma}g^{\gamma_1}_{\beta_1}\cdots g^{\gamma_{N-1})}_{\beta_{N-1}}-
g^{(\gamma}_{(\beta}g^{\gamma_1\gamma_2}\cdots g^{\gamma_{N-2}\gamma_{N-1})}
g_{\beta_1\beta_2}\cdots g_{\beta_{N-2}\beta_{N-1})}8\frac{N!!}{(N+3)!!}\right)+\\
 &+& \frac{\partial C_{h,k}^{\alpha
A_1\cdots A_h B_1\cdots B_k}} {\partial \mu_{\beta}}
g^{(\gamma}_{\beta}g^{\gamma_1\gamma_2}\cdots
g^{\gamma_{N-2}\gamma_{N-1})}8\frac{N!!}{(N+3)!!}(-m^2)^{\frac{N-1}{2}},
\end{eqnarray*}
where $(\ref{8b})_1$ and $(\ref{6})_2$ have been used.\\
Multiplying both sides by $g_{\gamma_1\gamma_2}\cdots g_{\gamma_{N-2}\gamma_{N-1}}$ we
obtain
\begin{equation}\label{9}
C_{h,k+1}^{\alpha A_1\cdots A_{h} B_1\cdots B_{k}\beta
\gamma_1\cdots \gamma_{N-1} }g_{\gamma_1\gamma_2}\cdots
g_{\gamma_{N-2}\gamma_{N-1}}=\frac{\partial C_{h,k}^{\alpha
A_1\cdots A_h B_1\cdots B_k}} {\partial \mu_{\beta}}
(-m^2)^{\frac{N-1}{2}}.
\end{equation}
So, from eq. $(\ref{5})$ we have obtained the compatibility conditions $(\ref{8})$ and $(\ref{9})$.\\
Let's proof the vice versa, i.e., that eq. $(\ref{5})$ is a
consequence of all the other equations. To this end it is firstly
useful to show a property of $\Delta h^{'\alpha}$. If we take its
derivative with respect to $\lambda_{\gamma_1\cdots \gamma_M}$ we
obtain, with passages like that used in eq. $(\ref{11a})$, that
\begin{eqnarray*}
\frac{\partial \Delta h^{'\alpha}}{\partial
\lambda_{\gamma_1\cdots
\gamma_M}}&=&\sum_{h,k=0}^{\infty}\frac{1}{h!k!}\frac{\partial
C_{h,k}^{\alpha A_1\cdots A_h B_1 \cdots B_k}}{\partial\lambda}
\frac{\partial \lambda}{\partial
\lambda_{\gamma_1\cdots\gamma_M}}\widetilde{\lambda}_{A_1}\cdots\widetilde{\lambda}_{A_h}\widetilde{\mu}_{B_1}\cdots
\widetilde{\mu}_{B_k}+\\
&+&\sum_{\stackrel {h,k=0}{h\neq
0}}^{\infty}\frac{h}{h!k!}C_{h,k}^{\alpha A_1\cdots A_h B_1\cdots
B_k}\widetilde{\lambda}_{A_1}\cdots\widetilde{\lambda}_{A_{h-1}}\widetilde{\mu}_{B_1}\cdots
\widetilde{\mu}_{B_k}\frac{\partial
\widetilde{\lambda}_{A_{h}}}{\partial
\lambda_{\gamma_1\cdots \gamma_M}}=\\
&=&\sum_{\stackrel {h,k=0}{h\neq
0}}^{\infty}\frac{h}{h!k!}C_{h,k}^{\alpha A_1\cdots A_h B_1\cdots
B_k}\widetilde{\lambda}_{A_1}\cdots\widetilde{\lambda}_{A_{h-1}}\widetilde{\mu}_{B_1}\cdots
\widetilde{\mu}_{B_k}g_{\alpha_1}^{(\gamma_1}\cdots
g_{\alpha_M}^{\gamma_M)}=\frac{\partial h^{'\alpha}}{\partial
\widetilde{\lambda}_{\gamma_1\cdots \gamma_M}}
\end{eqnarray*}
where we have used eq. $(\ref{8})$. So we have proved that derivation of eq. $(\ref{4})$
with respect to $\lambda_{\gamma_1\cdots \gamma_M}$ is equivalent to its derivation with
respect to $\widetilde{\lambda}_{\gamma_1\cdots \gamma_M}$.\\
With analogous passages, but by using eq. $(\ref{9})$, it is possible to prove that
derivation of eq. $(\ref{4})$ with respect to $\mu_{\gamma_1\cdots \gamma_N}$ is
equivalent to its
derivation with respect to $\widetilde{\mu}_{\gamma_1\cdots \gamma_N}$.\\
We are now ready to prove that eq. $(\ref{5})$ is a consequence of the other equations.
To this end, let us take the $h^{th}$ derivative of eq. $(\ref{4})$ with respect to
$\lambda_{\gamma_1\cdots \gamma_M}$, then its $k^{th}$ derivative with respect to
$\mu_{\gamma_1\cdots \gamma_N}$ and calculate the result at equilibrium; by using eqs.
$(\ref{8})$, $(\ref{9})$ and the above mentioned property , we obtain eq. $(\ref{5})$. In
other words
we can forget eq. $(\ref{5})$ and retain only eqs. $(\ref{8})$ and $(\ref{9})$.\\
So it remains to solve eqs. $(\ref{8})$ and $(\ref{9})$ in the unknown symmetric tensors
$C_{h,k}^{\cdots}$.\\
We notice that both sides in equation $(\ref{8})$ and the
left-hand side of eq. $(\ref{9})$ are symmetric; then we have to
impose that the right hand side of this last one is also
symmetric. In other words, both the tensors $C_{h,k}^{\cdots}$ and
their derivatives with respect to $\mu_{\beta}$ are symmetric. The
tensors satisfying this property are elements of a family
$\mathcal{F}$ which will be characterized in the following
section. Moreover, interesting properties of this family
$\mathcal{F}$ will be shown and they are useful to exploit our
eqs. $(\ref{8})$ and $(\ref{9})$. The effective exploitation will
be accomplished in section 5 for the case $N=1$ and in section 6
for the case $N>1$. We complete the present section simply
reporting the results, so that they are available to
whom is not interested in the proofs.\\
For the case $N>1$, they are
\begin{equation}\label{r5-1.1}
C_{h,k}^{\alpha_1\cdots
\alpha_{Mh+kN+1}}=\sum_{s=0}^{\left[\frac{Mh+Nk+1}{2}\right]}C^{h,k}_s
g^{(\alpha_1\alpha_2}\cdots
g^{\alpha_{2s-1}\alpha_{2s}}\mu^{\alpha_{2s+1}}\cdots
\mu^{\alpha_{Mh+Nk+1})},
\end{equation}
with
\begin{eqnarray}\label{r5-1.2}
C^{h,k}_s&=&2^{\stackrel{2\left[\frac{Mh+Nk+1}{2}\right]+\left[\frac{k}{2}\right]-2s}{~}}\cdot\frac{\left[\frac{Mh+Nk+1}{2}\right]!}{s!}
\frac{1}{(Mh+Nk+1-2s)!}\cdot\nonumber \\
&\cdot
&\gamma^{-6-Mh-(N+1)k+2s}\cdot\sum_{q=0}^{\frac{Mh+k(N-1)-2}{2}}\frac{\left[Mh+k(N-1)+1+2\left[\frac{k}{2}\right]\right]!!}{\left[hM+k(N-1)-2q-2\right]!!}\cdot\nonumber\\
&\cdot & (-m^2)^{\frac{N-1}{2}k+\frac{M}{2}h}\frac{d^h c_{k+Mh,q}}{d \lambda^h}
\cdot\frac{\left(q+2+\frac{Mh+(N+1)k}{2}-s\right)!}{(q+2)!}\cdot\left(
\frac{1}{\gamma^2}\right)^q,
\end{eqnarray}
and $c_{k,q}=c_{k,q}(\lambda)$ restricted only by $c_{k,q}=c_{k-1,q}$ for $q=0,\cdots,
\frac{[Mh+(k-1)(N-1)-2]}{2}$.\\
For the case $N=1$, and consequently k=0, the results are
\begin{equation}\label{r5-1.3}
C_{h}^{\alpha_1\cdots
\alpha_{Mh+1}}=\sum_{s=0}^{\frac{Mh}{2}}C^{h}_s
g^{(\alpha_1\alpha_2}\cdots
g^{\alpha_{2s-1}\alpha_{2s}}\mu^{\alpha_{2s+1}}\cdots
\mu^{\alpha_{Mh+1})},
\end{equation}
with
\begin{eqnarray}\label{r5-1.4}
C^{h}_s&=&2^{Mh-2s}\frac{\left(\frac{Mh}{2}\right)!}{s!}
\frac{1}{(Mh+1-2s)!}\cdot\gamma^{-6-Mh+2s}\cdot\sum_{q=0}^{\frac{Mh-2}{2}}\frac{\left(Mh+1\right)!!}{\left(Mh-2q-2\right)!!}\cdot\nonumber\\
&\cdot & (-m^2)^{\frac{M}{2}h}\frac{d^h c_{h,q}}{d \lambda^h}
\cdot\frac{\left(q+2+\frac{Mh}{2}-s\right)!}{(q+2)!}\cdot\left(
\frac{1}{\gamma^2}\right)^q,
\end{eqnarray}
and $c_{h,q}=c_{h,q}(\lambda)$ restricted only by $c_{h,q}=c_{h-1,q}$ for $q=0,\cdots,
\frac{[M(h-1)-2]}{2}$.
\section{On the family $\mathcal{F}$ and its properties}
Let us characterize now the family $\mathcal{F}$ of tensorial functions of a scalar
$\lambda$ and of a time-like 4-vector $\mu_{\beta}$, which are symmetric together with
their derivative with respect to $\mu_{\beta}$.\\
We will prove now that
\begin{proposition}\label{prop1}
Each element of the family  $\mathcal{F}$ can be written as
\begin{equation}\label{°°°}
\phi^{\alpha_1\cdots\alpha_n}(\lambda,\mu_{\beta})=\sum_{s=0}^{\left[\frac{n}{2}\right]}\phi_s^n(\lambda,\gamma)
g^{(\alpha_1\alpha_2}\cdots g^{\alpha_{2s-1}\alpha_{2s}}\mu^{\alpha_{2s+1}}\cdots
\mu^{\alpha_n)}
\end{equation}
with the scalars $\phi^n_s$satisfying the condition
\begin{equation}\label{triangolo}
\frac{\partial \phi_s^n}{\partial\gamma}\frac{1}{\gamma}2s+(n-2s+2)(n-2s+1)\phi_{s-1}^n=0
\quad \text{for }s=1,\cdots \left[\frac{n}{2}\right].
\end{equation}
(Note that among the terms $\phi_s^n$ we may call \underline{leading term} the one with
the highest value of s, i.e. $\phi_{\left[\frac{n}{2}\right]}^{n}$. Once the leading term
is known, we can find all the other terms present in eq. $(\ref{°°°})$ thanks to eq.
$(\ref{triangolo})$).
\end{proposition}
\begin{proof}
From the representation theorems we know that eq. $(\ref{°°°})$ is the most general
expression of a symmetric tensorial function depending on $\lambda$ and $\mu_{\beta}$.
Taking into account that $\frac{\partial \gamma}{\partial
\mu_{\beta}}=-\frac{\mu^{\beta}}{\gamma}$, the derivative of
$\phi^{\alpha_1\cdots\alpha_n}(\lambda,\mu_{\beta})$ with respect to $\mu_{\beta}$ is
\begin{eqnarray}\label{+}
\frac{\partial \phi^{\alpha_1\cdots\alpha_n}}{\partial \mu_{\beta}}&=&
\sum_{s=0}^{\left[\frac{n}{2}\right]}-\frac{\partial
\phi_s^n}{\partial\gamma}\frac{\mu^{\beta}}{\gamma} g^{(\alpha_1\alpha_2}\cdots
g^{\alpha_{2s-1}\alpha_{2s}}\mu^{\alpha_{2s+1}}\cdots \mu^{\alpha_n)}+\nonumber\\
&+& \sum_{s=0}^{\left[\frac{n}{2}\right]}\phi_s^n g^{(\alpha_1\alpha_2}\cdots
g^{\alpha_{2s-1}\alpha_{2s}}\mu^{\alpha_{2s+1}}\cdots
\mu^{\alpha_{n-1}}g^{\alpha_n)\beta}(n-2s).
\end{eqnarray}
To be symmetric, the expression above must be equal to its symmetric part with respect to
$\alpha_1\cdots\alpha_n\beta$, i.e. to
\begin{eqnarray*}
&& \sum_{s=0}^{\left[\frac{n}{2}\right]}-\frac{\partial
\phi_s^n}{\partial\gamma}\frac{1}{\gamma} g^{(\alpha_1\alpha_2}\cdots
g^{\alpha_{2s-1}\alpha_{2s}}\mu^{\alpha_{2s+1}}\cdots \mu^{\alpha_n}\mu^{\beta)}+\\
&+&
\sum_{S=1}^{\left[\frac{n}{2}\right]+1}\phi_{\stackrel{~}{S-1}}^n
g^{(\alpha_1\alpha_2}\cdots
g^{\alpha_{2S-1}\alpha_{2S}}\mu^{\alpha_{2S+1}}\cdots
\mu^{\alpha_n}\mu^{\beta)}(n-2S+2)=\\
&=&-\frac{\partial \phi_0^n}{\partial\gamma}\frac{1}{\gamma}\mu^{\alpha_{1}}\cdots
\mu^{\alpha_n}\mu^{\beta}+\phi_{\left[\frac{n}{2}\right]}^n
\left(n-2\left[\frac{n}{2}\right]\right)g^{(\alpha_1\alpha_2}\cdots g^{\alpha_n\beta)}+\\
&+&\sum_{s=1}^{\left[\frac{n}{2}\right]}\left[-\frac{\partial
\phi_s^n}{\partial\gamma}\frac{1}{\gamma}+ (n-2s+2)\phi_{s-1}^n\right]
g^{(\alpha_1\alpha_2}\cdots g^{\alpha_{2s-1}\alpha_{2s}}\mu^{\alpha_{2s+1}}\cdots
\mu^{\alpha_n}\mu^{\beta)}=\\
&=&-\frac{\partial \phi_0^n}{\partial\gamma}\frac{1}{\gamma}\mu^{\alpha_{1}}\cdots
\mu^{\alpha_n}\mu^{\beta}+\phi_{\left[\frac{n}{2}\right]}^n
\left(n-2\left[\frac{n}{2}\right]\right)g^{(\alpha_1\alpha_2}\cdots g^{\alpha_n\beta)}+\\
&+&\sum_{s=1}^{\left[\frac{n}{2}\right]}\Bigg[-\frac{\partial
\phi_s^n}{\partial\gamma}\frac{1}{\gamma}+ (n-2s+2)\phi_{s-1}^n\Bigg]
\Bigg[\frac{2s}{n+1}g^{\beta(\alpha_1}\cdots
g^{\alpha_{2s-2}\alpha_{2s-1}}\mu^{\alpha_{2s}}\cdots
\mu^{\alpha_n)}+\\
&+& \frac{n+1-2s}{n+1}g^{(\alpha_1\alpha_2}\cdots
g^{\alpha_{2s-1}\alpha_{2s}}\mu^{\alpha_{2s+1}}\cdots \mu^{\alpha_n)}\mu^{\beta}\Bigg],
\end{eqnarray*}
where in the second row we have substituted $s=S-1$. In the third row we have reported
the term coming from the first row with s=0, and that from the second row with
$S=\left[\frac{n}{2}\right]+1$; in the fourth row the remaining terms from first and
second row.\\
Comparing the result with eq. $(\ref{+})$, we obtain
\begin{eqnarray*}
&&\sum_{s=0}^{\left[\frac{n}{2}\right]}-\frac{\partial
\phi_s^n}{\partial\gamma}\frac{1}{\gamma} g^{(\alpha_1\alpha_2}\cdots
g^{\alpha_{2s-1}\alpha_{2s}}\mu^{\alpha_{2s+1}}\cdots \mu^{\alpha_n)}=\\
&=&-\frac{\partial \phi_0^n}{\partial\gamma}\frac{1}{\gamma}\mu^{\alpha_{1}}\cdots
\mu^{\alpha_n}+\\
&+&\sum_{s=1}^{\left[\frac{n}{2}\right]}\Bigg[-\frac{\partial
\phi_s^n}{\partial\gamma}\frac{1}{\gamma}+
(n-2s+2)\phi_{s-1}^n\Bigg]\frac{n+1-2s}{n+1}g^{(\alpha_1\alpha_2}\cdots
g^{\alpha_{2s-1}\alpha_{2s}}\mu^{\alpha_{2s+1}}\cdots \mu^{\alpha_n)}
\end{eqnarray*}
and
\begin{eqnarray*}
&&\sum_{s=0}^{\left[\frac{n}{2}\right]}\phi_s^n (n-2s) g^{\beta(\alpha_1}\cdots
g^{\alpha_{2s}\alpha_{2s+1}}\mu^{\alpha_{2s+2}}\cdots \mu^{\alpha_{n})}=\\
&=&\phi_{\left[\frac{n}{2}\right]}^n
\left(n-2\left[\frac{n}{2}\right]\right)g^{(\alpha_1\alpha_2}\cdots g^{\alpha_n)\beta}+\\
&+&\sum_{s=0}^{\left[\frac{n}{2}\right]-1}\Bigg[-\frac{\partial
\phi_{s+1}^n}{\partial\gamma}\frac{1}{\gamma}+ (n-2s)\phi_{s}^n\Bigg]
\frac{2s+2}{n+1}g^{\beta(\alpha_1}\cdots
g^{\alpha_{2s}\alpha_{2s+1}}\mu^{\alpha_{2s+2}}\cdots \mu^{\alpha_n)}
\end{eqnarray*}
where we have considered $s=S+1$ and then $S=s$ in the last term.\\
So we have
\begin{equation*}
\frac{\partial \phi_s^n}{\partial\gamma}\frac{1}{\gamma}2s+(n-2s+2)(n-2s+1)\phi_{s-1}^n=0
\quad \text{for }s=1,\cdots \left[\frac{n}{2}\right]
\end{equation*}
and
\begin{equation*}
\frac{\partial
\phi_{s+1}^n}{\partial\gamma}\frac{1}{\gamma}(2s+2)+(n-2s)(n-2s-1)\phi_{s}^n=0 \quad
\text{for }s=0,\cdots \left[\frac{n-2}{2}\right].
\end{equation*}
We can see that the second of these equations coincides with the first one. So the
characteristic condition for $\mathcal{F}$ is the above reported eq.
$(\ref{triangolo})$.\\
This completes the proof of Proposition \ref{prop1}.
\end{proof}\\
It will be useful for the sequel to note that, thanks to $(\ref{triangolo})$, eq.
$(\ref{+})$ becomes
\begin{eqnarray}\label{triangolo3}
\frac{\partial \phi^{\alpha_1\cdots\alpha_n}}{\partial \mu_{\beta}}=
\sum_{s=0}^{\left[\frac{n+1}{2}\right]} \phi_s^{n+1}(\lambda,\gamma)
g^{(\alpha_1\alpha_2}\cdots g^{\alpha_{2s-1}\alpha_{2s}}\mu^{\alpha_{2s+1}}\cdots
\mu^{\alpha_n}\mu^{\beta)}
\end{eqnarray}
with
\begin{eqnarray}\label{triangolo2}
  \begin{cases}
    \phi_0^{n+1}=-\frac{1}{\gamma}\frac{\partial \phi_0^n}{\partial\gamma}, & \\
    \phi_s^{n+1}=\frac{n+1}{2s}(n-2s+2) \phi_{s-1}^n & \text{for }s=1,\cdots,\left[\frac{n+1}{2}\right].
  \end{cases}
\end{eqnarray}
This allows to prove the following
\begin{proposition}\label{prop2}
If $\phi^{\alpha_1\cdots\alpha_n}\in\mathcal{F}$, it follows that also $\frac{\partial
\phi^{\alpha_1\cdots\alpha_n}}{\partial\mu_{\beta}}\in\mathcal{F}$.
\end{proposition}
\begin{proof}
Because of eq. $(\ref{triangolo3})$, to demonstrate the theorem it will be sufficient to
prove that eq. $(\ref{triangolo})$ holds true also with n+1 instead of n, i.e.
\begin{equation}\label{33}
\frac{\partial
\phi_s^{n+1}}{\partial\gamma}\frac{1}{\gamma}2s+(n-2s+3)(n-2s+2)\phi_{s-1}^{n+1}=0 \quad
\text{for }s=1,\cdots \left[\frac{n+1}{2}\right].
\end{equation}
For s=1, thanks to $(\ref{triangolo2})$, it becomes
\begin{equation*}
\frac{\partial
\phi_0^{n}}{\partial\gamma}\frac{2}{\gamma}\frac{n}{2}(n+1)+(n+1)n\left(-\frac{1}{\gamma}\right)\frac{\phi_{0}^{n}}{\partial\gamma}=0
\end{equation*}
that is identically satisfied.\\
Instead, for $s=2,\cdots,\left[\frac{n+1}{2}\right]$, thanks to $(\ref{triangolo2})_2$,
it becomes
\begin{equation*}
\frac{\partial
\phi_{s-1}^{n}}{\partial\gamma}\frac{2s}{\gamma}\frac{n+1}{2s}(n-2s+2)+(n-2s+3)(n-2s+2)\frac{n+1}{2s-2}(n-2s+4)\phi_{s-2}^{n}=0
\end{equation*}
that is eq. $(\ref{triangolo})$ with s-1 instead of s, after having divided it for 2s-2.
\end{proof}\\
It is also important the following
\begin{proposition}\label{prop3}
If $\phi^{\alpha_1\cdots\alpha_{n+1}}\in\mathcal{F}$, then
$\phi^{\alpha_1\cdots\alpha_{n}}\in\mathcal{F}$ exists such that
$\phi^{\alpha_1\cdots\alpha_{n+1}}=\frac{\partial
\phi^{\alpha_1\cdots\alpha_n}}{\partial\mu_{\alpha_{n+1}}}$.
\end{proposition}
\begin{proof}
In fact the integrability condition for the problem
\begin{equation*}
\phi^{\alpha_1\cdots\alpha_{n+1}}=\frac{\partial
\phi^{\alpha_1\cdots\alpha_n}}{\partial\mu_{\alpha_{n+1}}}
\end{equation*}
is
\begin{equation*}
\frac{\partial \phi^{\alpha_1\cdots\alpha_{n-1}[\alpha_n}}{\partial\mu_{\alpha_{n+1}]}}=0
\end{equation*}
which is surely satisfied because $\phi^{\alpha_1\cdots\alpha_{n}}\in\mathcal{F}$ and,
consequently, it and its derivative with respect to $\mu_{\alpha_{n+1}}$ are symmetric.\\
We note also that, if $\widetilde{\phi}^{\alpha_1\cdots\alpha_{n}}$ is a particular
solution of this problem, then the general one is
\begin{equation*}
\phi^{\alpha_1\cdots\alpha_{n}}=
  \begin{cases}
    \widetilde{\phi}^{\alpha_1\cdots\alpha_{n}}+\phi(\lambda)g^{(\alpha_1\alpha_2}\cdots g^{\alpha_{n-1}\alpha_n)} & \text{if n is even}, \\
    \widetilde{\phi}^{\alpha_1\cdots\alpha_{n}} & \text{if n is odd},
  \end{cases}
\end{equation*}
with $\phi(\lambda)$ a scalar function.
\end{proof}
\begin{proposition}\label{prop4}
If $\phi^{\alpha_1\cdots\alpha_{n+r}}\in\mathcal{F}$, then
$\phi^{\alpha_1\cdots\alpha_{n}}\in\mathcal{F}$ exists such that
$\phi^{\alpha_1\cdots\alpha_{n+r}}=\frac{\partial^r
\phi^{\alpha_1\cdots\alpha_n}}{\partial\mu_{\alpha_{n+1}}\cdots\mu_{\alpha_{n+r}}}$.\\
Moreover, if $\widetilde{\phi}^{\alpha_1\cdots\alpha_{n}}$ is a particular solution of
this problem, then the general one is
\begin{equation*}
\phi^{\alpha_1\cdots\alpha_{n}}=
\begin{cases}
\widetilde{\phi}^{\alpha_1\cdots\alpha_{n}}+\sum_{i=0}^{\left[\frac{r-1}{2}\right]}\phi_i(\lambda)
g^{(\alpha_1\alpha_2}\cdots g^{\alpha_{n+2i-1}\alpha_n+2i)}\mu_{\alpha_{n+1}}\cdots\mu_{\alpha_{n+2i}} & \text{if n is even}, \\
~\\
\widetilde{\phi}^{\alpha_1\cdots\alpha_{n}}+\sum_{i=0}^{\left[\frac{r-2}{2}\right]}\phi_i(\lambda)
g^{(\alpha_1\alpha_2}\cdots
g^{\alpha_{n+2i}\alpha_n+2i+1)}\mu_{\alpha_{n+1}}\cdots\mu_{\alpha_{n+2i+1}}
& \text{if n is odd},
\end{cases}
\end{equation*}
with $\phi_i(\lambda)$ scalar functions.
\end{proposition}
\begin{proof}
We can prove this proposition with the iterative procedure.\\
It holds for r=1 for the Proposition \ref{prop3}.\\
Let us assume now that it holds for $r=\overline{r}$ and prove that it is satisfied also
when $r=\overline{r}+1$.\\
If $\phi^{\alpha_1\cdots\alpha_{n+\overline{r}+1}}\in\mathcal{F}$ we can apply this
proposition with n+1 instead of n and $\overline{r}$ instead of r. Then we have that
$\phi^{\alpha_1\cdots\alpha_{n+1}}\in \mathcal{F}$ exists such that
\begin{equation}\label{xx}
\phi^{\alpha_1\cdots\alpha_{n+\overline{r}+1}}=\frac{\partial^{\overline{r}}\phi^{\alpha_1\cdots\alpha_{n+1}}}{\partial\mu_{\alpha_{n+2}}\cdots\partial\mu_{\alpha_{n+\overline{r}+1}}}.
\end{equation}
But, for the first part of Proposition \ref{prop3} we have that
$\phi^{\alpha_1\cdots\alpha_{n}}\in\mathcal{F}$ exists such that
$\phi^{\alpha_1\cdots\alpha_{n+1}}=\frac{\partial\phi^{\alpha_1\cdots\alpha_{n}}}{\partial\mu_{\alpha_{n+1}}}$
which, substituted in $(\ref{xx})$ gives
\begin{equation}\label{xy}
\phi^{\alpha_1\cdots\alpha_{n+\overline{r}+1}}=\frac{\partial^{\overline{r}+1}\phi^{\alpha_1\cdots\alpha_{n}}}{\partial\mu_{\alpha_{n+1}}\cdots\partial\mu_{\alpha_{n+\overline{r}+1}}}.
\end{equation}
So the existence of solutions has been proved.\\
If $\widetilde{\phi}^{\alpha_1\cdots\alpha_{n}}$ is a particular of these solutions, we
have
\begin{equation*}
\phi^{\alpha_1\cdots\alpha_{n+\overline{r}+1}}=\frac{\partial^{\overline{r}+1}\widetilde{\phi}^{\alpha_1\cdots\alpha_{n}}}{\partial\mu_{\alpha_{n+1}}\cdots\partial\mu_{\alpha_{n+\overline{r}+1}}},
\end{equation*}
which, together with eq. $(\ref{xy})$ implies
\begin{equation*}
\frac{\partial^{\overline{r}}}{\partial\mu_{\alpha_{n+1}}\cdots\partial\mu_{\alpha_{n+\overline{r}}}}
\left[\frac{\partial}{\partial\mu_{\alpha_{n+\overline{r}+1}}}\left(\phi^{\alpha_1\cdots\alpha_{n}}-\widetilde{\phi}^{\alpha_1\cdots\alpha_{n}}\right)\right]=0.
\end{equation*}
By applying the second part of this Proposition \ref{prop4}, we conclude that
\begin{equation*}
\frac{\partial}{\partial\mu_{\alpha_{n+\overline{r}+1}}}\left(\phi^{\alpha_1\cdots\alpha_{n}}-\widetilde{\phi}^{\alpha_1\cdots\alpha_{n}}\right)=
  \begin{cases}
    \sum_{i=0}^{\left[\frac{\overline{r}-1}{2}\right]}\overline{\phi}_i(\lambda)g^{(\alpha_1\alpha_2}\cdots g^{\alpha_{n+2i}\alpha_{n+2i+1})}\mu_{\alpha_{n+2}}\cdots\mu_{\alpha_{n+2i+1}} \\
    \text{if n+1 is even}, \\
    ~\\
    \sum_{i=0}^{\left[\frac{\overline{r}-2}{2}\right]}\overline{\phi}_i(\lambda)g^{(\alpha_1\alpha_2}\cdots g^{\alpha_{n+2i+1}\alpha_{n+2i+2})}\mu_{\alpha_{n+2}}\cdots\mu_{\alpha_{n+2i+2}} \\
    \text{if n+1 is odd}.
  \end{cases}
\end{equation*}
This relation can be integrated and gives
\begin{equation*}
\phi^{\alpha_1\cdots\alpha_{n}}-\widetilde{\phi}^{\alpha_1\cdots\alpha_{n}}=
  \begin{cases}
    \sum_{i=0}^{\left[\frac{\overline{r}-1}{2}\right]}\overline{\phi}_i(\lambda)\frac{1}{2i+1}g^{(\alpha_1\alpha_2}\cdots g^{\alpha_{n+2i}\alpha_{n+2i+1})}\mu_{\alpha_{n+1}}\cdots\mu_{\alpha_{n+2i+1}} \\
    \qquad \qquad \qquad \qquad \qquad \qquad\text{if n is odd}, \\
    ~
    \sum_{i=0}^{\left[\frac{\overline{r}-2}{2}\right]}\overline{\phi}_i(\lambda)\frac{1}{2i+2}g^{(\alpha_1\alpha_2}\cdots g^{\alpha_{n+2i+1}\alpha_{n+2i+2})}\mu_{\alpha_{n+1}}\cdots\mu_{\alpha_{n+2i+2}} + \\
    + \phi_0(\lambda)g^{(\alpha_1\alpha_2}\cdots g^{\alpha_{n-1}\alpha_{n})}  \quad \text{if n is even}.
  \end{cases}
\end{equation*}
So we have obtained the second part of this proposition, with $\overline{r}+1$ instead of
$r$, and
\begin{equation*}
\phi_i(\lambda)=
  \begin{cases}
    \overline{\phi}_i(\lambda)\frac{1}{2i+1} & \text{if n is odd}, \\
    \overline{\phi}_{i-1}(\lambda)\frac{1}{2i} & \text{if n is even (For this case we have put i=I-1)}.
  \end{cases}
\end{equation*}
This completes the proof of Proposition \ref{prop4}.
\end{proof}
\begin{proposition}\label{prop5}
If $\phi^{\alpha_1\cdots\alpha_{n}}\in \mathcal{F}$, then
$\phi^{\alpha_1\cdots\alpha_{n+2}}\in \mathcal{F}$ exists such that
\begin{equation}\label{tb}
\phi^{\alpha_1\cdots\alpha_{n+2}}g_{\alpha_{n+1}\alpha_{n+2}}=\phi^{\alpha_1\cdots\alpha_{n}}.
\end{equation}
Moreover, its leading term is
\begin{equation}\label{36a}
\phi^{n+2}_{\left[\frac{n+2}{2}\right]}=\gamma^{-2\left(n+3-\left[\frac{n+2}{2}\right]\right)}
\frac{(n+1)(n+2)}{2\left[\frac{n+2}{2}\right]}\left[\int\phi^n_{\left[\frac{n}{2}\right]}
\gamma^{2\left(n+3-\left[\frac{n+2}{2}\right]\right)-1}
d\gamma+f^{n+2}_{\left[\frac{n+2}{2}\right]}(\lambda)\right].
\end{equation}
with $\phi_{\left[\frac{n}{2}\right]}^n$ leading term of
$\phi^{\alpha_1\cdots\alpha_{n}}$ and
$f^{n+2}_{\left[\frac{n+2}{2}\right]}$ an arbitrary function.
\end{proposition}
\begin{proof}
Because of $\phi^{\alpha_1\cdots\alpha_{n+2}}\in\mathcal{F}$, it has the form
$(\ref{°°°})$ with coefficients satisfying eq. $(\ref{triangolo})$ and n+2 instead of n,
i.e.,
\begin{equation*}
\phi^{\alpha_1\cdots\alpha_{n+2}}(\lambda,\mu_{\beta})=\sum_{s=0}^{\left[\frac{n+2}{2}\right]}\phi_s^{n+2}(\lambda,\gamma)
g^{(\alpha_1\alpha_2}\cdots g^{\alpha_{2s-1}\alpha_{2s}}\mu^{\alpha_{2s+1}}\cdots
\mu^{\alpha_{n+2)}}
\end{equation*}
with
\begin{equation}\label{triangolobis}
\frac{\partial
\phi_s^{n+2}}{\partial\gamma}\frac{1}{\gamma}2s+(n-2s+4)(n-2s+3)\phi_{s-1}^{n+2}=0 \quad
\text{for }s=1,\cdots \left[\frac{n+2}{2}\right].
\end{equation}
Let's substitute this expression of $\phi^{\alpha_1\cdots\alpha_{n+2}}$ in the left hand
side of eq. $(\ref{tb})$ and explicit the symmetrization, so obtaining
\begin{eqnarray*}
\phi^{\alpha_1\cdots\alpha_{n+2}}g_{\alpha_{n+1}\alpha_{n+2}}&=&\sum_{s=0}^{\left[\frac{n+2}{2}\right]}\phi_s^{n+2}(\lambda,\gamma)
g^{(\alpha_1\alpha_2}\cdots g^{\alpha_{2s-1}\alpha_{2s}}\mu^{\alpha_{2s+1}}\cdots
\mu^{\alpha_{n+2)}}g_{\alpha_{n+1}\alpha_{n+2}}= \\
 &=&\sum_{s=1}^{\left[\frac{n+2}{2}\right]}\phi_s^{n+2}(\lambda,\gamma)
\frac{2s(2s+2)}{(n+2)(n+1)}g^{(\alpha_1\alpha_2}\cdots
g^{\alpha_{2s-3}\alpha_{2s-2}}\mu^{\alpha_{2s-1}}\cdots
\mu^{\alpha_{n)}}+\\
&+&\sum_{s=1}^{\left[\frac{n+2}{2}\right]}\phi_s^{n+2}(\lambda,\gamma)
2\frac{2s(n-2s+2)}{(n+2)(n+1)}g^{(\alpha_1\alpha_2}\cdots
g^{\alpha_{2s-3}\alpha_{2s-2}}\mu^{\alpha_{2s-1}}\cdots \mu^{\alpha_{n)}}+\\
&+&\sum_{s=0}^{\left[\frac{n}{2}\right]}\phi_s^{n+2}(\lambda,\gamma)
\frac{(n+2-2s)(n+2-2s-1)}{(n+2)(n+1)}\cdot\\
&&g^{(\alpha_1\alpha_2}\cdots g^{\alpha_{2s-1}\alpha_{2s}}\mu^{\alpha_{2s+1}}\cdots
\mu^{\alpha_{n)}}(-\gamma^2).
\end{eqnarray*}
Blending the first two sums and putting S=s-1 we obtain
\begin{eqnarray*}
\phi^{\alpha_1\cdots\alpha_{n+2}}g_{\alpha_{n+1}\alpha_{n+2}}&=&\sum_{s=0}^{\left[\frac{n}{2}\right]}
\left\{\phi_{s+1}^{n+2}4\frac{(s+1)(n-s+2)}{(n+2)(n+1)}+\phi_s^{n+2}(-\gamma^2)
\frac{(n+2-2s)(n+1-2s)}{(n+2)(n+1)}\right\} \\
&&g^{(\alpha_1\alpha_2}\cdots g^{\alpha_{2s-1}\alpha_{2s}}\mu^{\alpha_{2s+1}}\cdots
\mu^{\alpha_{n)}}.
\end{eqnarray*}
With the use of this expression and of eq. $(\ref{°°°})$ for
$\phi^{\alpha_1\cdots\alpha_{n}}$, eq. $(\ref{tb})$ becomes
\begin{eqnarray*}
&&\sum_{s=0}^{\left[\frac{n}{2}\right]}
\left\{\phi_{s+1}^{n+2}4\frac{(s+1)(n-s+2)}{(n+2)(n+1)}+\phi_s^{n+2}(-\gamma^2)
\frac{(n+2-2s)(n+1-2s)}{(n+2)(n+1)}\right\}\cdot \\
&&g^{(\alpha_1\alpha_2}\cdots g^{\alpha_{2s-1}\alpha_{2s}}\mu^{\alpha_{2s+1}}\cdots
\mu^{\alpha_{n)}}= \sum_{s=0}^{\left[\frac{n}{2}\right]}
\phi_s^{n}g^{(\alpha_1\alpha_2}\cdots
g^{\alpha_{2s-1}\alpha_{2s}}\mu^{\alpha_{2s+1}}\cdots \mu^{\alpha_{n)}}.
\end{eqnarray*}
i.e.
\begin{eqnarray}\label{*}
 \phi_s^{n}=\frac{1}{(n+2)(n+1)}\left[\phi_{s+1}^{n+2}4(s+1)(n-s+2)+\phi_s^{n+2}(-\gamma^2)
(n+2-2s)(n+1-2s)\right],
\end{eqnarray}
for $s=0,\cdots,\left[\frac{n}{2}\right]$.\\
If we use eqs. $(\ref{triangolo})$ and $(\ref{triangolobis})$, this expression becomes
\begin{eqnarray}\label{**}
\frac{\partial\phi_{s+1}^{n}}{\partial\gamma}=\frac{1}{(n+2)(n+1)}
\left[\frac{\partial\phi_{s+2}^{n+2}}{\partial\gamma}4(s+2)(n-s+2)+
\frac{\partial\phi_{s+1}^{n+2}}{\partial\gamma}(-\gamma^2) (n-2s)(n-1-2s)\right]
\end{eqnarray}
that is eq. $(\ref{*})$ with s+1 instead of s, derivated with respect to $\gamma$ and
with another use of eq. $(\ref{triangolobis})$. So we have proved that if eq. $(\ref{*})$
holds true for a particular value of s, it will hold true also for all lower values of s;
so it suffices to impose eq. $(\ref{*})$ for the bigger value of s, i.e.,
\begin{equation}\label{38a}
\phi_{\left[\frac{n}{2}\right]}^{n}=\frac{2\left[\frac{n}{2}\right]+2}{(n+2)(n+1)}
\left[2\phi_{\left[\frac{n}{2}\right]}^{n+2}\left(n-\left[\frac{n}{2}\right]+2\right)
+\gamma\frac{\partial}{\partial\gamma}\phi_{\left[\frac{n+2}{2}\right]}^{n+2} \right],
\end{equation}
where eq. $(\ref{triangolobis})$ has been used. The general solution of this equation is
reported in eq. $(\ref{36a})$.
\end{proof}\\
It will be useful also the following
\begin{proposition}\label{prop6}
If the leading term of $\phi^{\alpha_1\cdots\alpha_{n}}$ is
$\phi^n_{\left[\frac{n}{2}\right]}=f(\lambda)\gamma^{-2(3+p)}$
with p a non negative integer, then the leading term of
$\phi^{\alpha_1\cdots\alpha_{n}}g_{\alpha_{n-2r+1}\alpha_{n-2r+2}}\cdots
g_{\alpha_{n-1}\alpha_n}$ is
\begin{equation}\label{eq*}
\phi^{n-2r}_{\left[\frac{n-2r}{2}\right]}=\frac{\left(2\left[\frac{n+1}{2}\right]-2r-1\right)!!}
{\left(2\left[\frac{n+1}{2}\right]-1\right)!!}f(\lambda)\gamma^{-2(3+p)}
\eta\left(2\left[\frac{n+1}{2}\right]-2r-2-2p,2\left[\frac{n+1}{2}\right]-4-2p\right)
\end{equation}
where, if $a\leq b$, $\eta(a,b)$ denotes the product of all even numbers between a and b,
while if $a=b+r$ then $\eta(a,b)=1$.
\end{proposition}
\begin{proof}
Let us prove eq. $(\ref{eq*})$ with the iterative procedure.\\
In the case r=0 it is an identity.\\
Let us assume that eq. $(\ref{eq*})$ holds up to the integer r and let us prove it with
r+1 instead of r.\\
Let us distinguish the cases with n odd and with n even; for the first one, by applying
eq. $(\ref{38a})$ with n-2r-2 instead of n, we obtain
\begin{eqnarray*}
&&\phi_{\frac{n-2r-3}{2}}^{n-2r-2}=\frac{1}{n-2r}
\left[(n-2r+3)\phi_{\frac{n-2r-1}{2}}^{n-2r}
+\gamma\frac{\partial}{\partial\gamma}\phi_{\frac{n-2r-1}{2}}^{n-2r}\right]=\nonumber \\
&& \frac{(n-2r-2)!!}{n!!}f(\lambda)\eta(n-1-2p-2r,n-3-2p)\gamma^{-2(3+p)}(n-2r+3-6-2p)
\end{eqnarray*}
that is eq. $(\ref{eq*})$ with r+1 instead of r.\\
In the case with n even, by applying eq. $(\ref{38a})$ with n-2r-2 instead of n, we
obtain
\begin{eqnarray*}
&&\phi_{\frac{n-2r-2}{2}}^{n-2r-2}=\frac{1}{n-2r-1}
\left[(n-2r+2)\phi_{\frac{n-2r}{2}}^{n-2r}
+\gamma\frac{\partial}{\partial\gamma}\phi_{\frac{n-2r}{2}}^{n-2r}\right]= \\
&&\frac{(n-2r-3)!!}{(n-1)!!}f(\lambda)\eta(n-2-2p-2r,n-4-2p)\gamma^{-2(3+p)}(n-2r-4-2p)
\end{eqnarray*}
that is eq. $(\ref{eq*})$ with r+1 instead of r.\\
\end{proof}
\begin{proposition}\label{prop7}
If $\phi^{\alpha_1\cdots\alpha_m}\in\mathcal{F}$ with leading term
$f(\lambda)\gamma^{-2(3+p)}$, with p a non negative integer such that
$p<m-\left[\frac{m}{2}\right]-1$ or $p>m-\left[\frac{m}{2}\right]+r-2$, then
$\phi^{\alpha_1\cdots\alpha_{m+2r}}\in\mathcal{F}$ exists such that
$\phi^{\alpha_1\cdots\alpha_m\alpha_{m+1}\alpha_{m+2}\cdots\alpha_{m+2r-1}\alpha_{m+2r}}g_{\alpha_{m+1}\alpha_{m+2}}
\cdots g_{\alpha_{m+2r-1}\alpha_{m+2r}}=\phi^{\alpha_1\cdots\alpha_m}$.\\
Moreover the leading term of $\phi^{\alpha_1\cdots\alpha_{m+2r}}$ is
\begin{equation}\label{39}
\frac{(m+2r)!}{m!}\frac{\left(2\left[\frac{m}{2}\right]\right)!!}{\left(2\left[\frac{m}{2}\right]+2r\right)!!}
\frac{\left(2m-2\left[\frac{m}{2}\right]-2p-4\right)!!}{\left(2m-2\left[\frac{m}{2}\right]+2r-2p-4\right)!!}
f(\lambda)\gamma^{-2(3+p)} +
\sum_{i=0}^{r-1}f_{i,r}(\lambda)\gamma^{-2\left(3+m+i-\left[\frac{m+2}{2}\right]\right)},
\end{equation}
with $f_{i,r}(\lambda)$ arbitrary functions.
\end{proposition}
\begin{proof}
Let us prove this proposition with the iterative procedure. \\
It is easy to verify that eq. $(\ref{39})$ holds for r=0.\\
Let us assume that it holds up to the integer r and prove that it holds also with r+1
instead of r.\\
Then we have to face the problem
\begin{equation}\label{40}
\phi^{\alpha_1\cdots\alpha_m\alpha_{m+1}\alpha_{m+2}\cdots\alpha_{m+2r+1}\alpha_{m+2r+2}}g_{\alpha_{m+1}\alpha_{m+2}}
\cdots g_{\alpha_{m+2r+1}\alpha_{m+2r+2}}=\phi^{\alpha_1\cdots\alpha_m}
\end{equation}
By defining
\begin{equation}\label{41}
\phi^{\alpha_1\cdots\alpha_{m+2r}}=\phi^{\alpha_1\cdots\alpha_{m+2r}\alpha_{m+2r+1}\alpha_{m+2r+2}}
g_{\alpha_{m+2r+1}\alpha_{m+2r+2}},
\end{equation}
the problem $(\ref{40})$ becomes that of the present proposition, which we have assumed
holding. Therefore, the leading term of $\phi^{\alpha_1\cdots\alpha_{m+2r}}$ is
$(\ref{39})$ and it remains to face only the problem $(\ref{41})$; by applying
Proposition \ref{prop5} with n=m+2r we find that the leading term of
$\phi^{\alpha_1\cdots\alpha_{m+2r+2}}$ is
\begin{eqnarray*}
&&\gamma^{-2\left(3+m+2r-\left[\frac{m+2r+2}{2}\right]\right)}
\frac{(m+2r+1)(m+2r+2)}{2\left[\frac{m+2r+2}{2}\right]}\Bigg[\int
\frac{(m+2r)!}{m!}\frac{\left(2\left[\frac{m}{2}\right]\right)!!}{\left(2\left[\frac{m}{2}\right]+2r\right)!!}\\
&&\frac{\left(2m-2\left[\frac{m}{2}\right]-2p-4\right)!!}{\left(2m-2\left[\frac{m}{2}\right]+2r-2p-4\right)!!}
f(\lambda)\gamma^{-2p+2m+4r-2\left[\frac{m+2r+2}{2}\right]-1} +
\sum_{i=0}^{r-1}f_{i,r}(\lambda)\gamma^{-2i+2r-1}d\gamma+f_{\left[\frac{m+2r+2}{2}\right]}^{m+2r+2}(\lambda)\Bigg]=\\
&&=\gamma^{-6} \frac{(m+2r+1)(m+2r+2)}{2\left[\frac{m+2r+2}{2}\right]}
\frac{(m+2r)!}{m!}\frac{\left(2\left[\frac{m}{2}\right]\right)!!}{\left(2\left[\frac{m}{2}\right]+2r\right)!!}
\frac{\left(2m-2\left[\frac{m}{2}\right]-2p-4\right)!!}{\left(2m-2\left[\frac{m}{2}\right]+2r-2p-4\right)!!}\cdot\\
&&\cdot f(\lambda)\frac{1}{-2p+2m+4r-2\left[\frac{m+2r+2}{2}\right]}\gamma^{-2p} +
\sum_{i=0}^{r-1}f_{i,r}(\lambda)
\frac{(m+2r+1)(m+2r+2)}{2\left[\frac{m+2r+2}{2}\right]}\cdot \\
&&\cdot \frac{1}{2r-2i}
\gamma^{-2i-2m-6+2\left[\frac{m+2}{2}\right]}+f_{\left[\frac{m+2r+2}{2}\right]}^{m+2r+2}(\lambda)
\frac{(m+2r+1)(m+2r+2)}{2\left[\frac{m+2r+2}{2}\right]}
\gamma^{-2\left(m+2r+3-\left[\frac{m+2r+2}{2}\right]\right)}
\end{eqnarray*}
that is eq. $(\ref{39})$ with r+1 instead of r and
\begin{equation*}
f_{i,r+1}=
  \begin{cases}
    f_{i,r}\frac{(m+2r+1)(m+2r+2)}{2\left[\frac{m+2r+2}{2}\right]}\frac{1}{2r-2i}& \text{for i=0, ... ,r-1}, \\
    f_{\left[\frac{m+2r+2}{2}\right]}^{m+2r+2}(\lambda)\frac{(m+2r+1)(m+2r+2)}{2\left[\frac{m+2r+2}{2}\right]}& \text{for i=r}.
  \end{cases}
\end{equation*}
This completes the proof.
\end{proof}\\
We conclude this section with the
\begin{proposition}\label{prop8}
If n is an even number, the tensor $g^{(\alpha_1\alpha_2}\cdots
g^{\alpha_{n-1}\alpha_n)}\mu_{\alpha_{n-r+1}}\cdots \mu_{\alpha_n}$ belongs to
$\mathcal{F}$ and, moreover,
\begin{equation}\label{15r-4.1}
g^{(\alpha_1\alpha_2}\cdots g^{\alpha_{n-1}\alpha_n)}\mu_{\alpha_{n-r+1}}\cdots
\mu_{\alpha_n}=\sum_{s=0}^{\left[\frac{n-r}{2}\right]}\phi_{s,r}^{n-r}g^{(\alpha_1\alpha_2}\cdots
g^{\alpha_{2s-1}\alpha_{2s}}\mu^{\alpha_{2s+1}}\cdots \mu^{\alpha_{n-r})}
\end{equation}
with
\begin{equation}\label{15r-4.2}
\phi_{s,r}^{n-r}=
  \begin{cases}
    1 & \text{for $r\leq 1$ and $s=\left[\frac{n-r}{2}\right]$}, \\
    0 & \text{for $r\leq 1$ and $0\leq s\leq\left[\frac{n-r}{2}\right]-1$}, \\
    \frac{r!(n-r)!}{(2s+2r-n)!!(2s)!!(n-r-2s)!(n-1)!!}\left(-\gamma^2\right)^{s+r-\frac{n}{2}} & \text{for $r\geq 2$ and $\frac{n}{2}-r\leq s\leq \left[\frac{n-r}{2}\right]$}, \\
    0 & \text{for $r\geq 2$ and $0\leq s\leq\frac{n}{2}-r-1$}.
  \end{cases}
\end{equation}
\end{proposition}
\begin{proof}
It is easy to note that the above tensor is an element of $\mathcal{F}$ because it and
its derivative with respect to $\mu_{\beta}$ are manifestly symmetric.\\
Let us prove $(\ref{15r-4.2})$ with the iterative procedure.\\
It is easy to verify that it holds when r=0.\\
When r=1 it is a consequence of
\begin{equation}\label{15r-4.3}
g^{(\alpha_1\alpha_2}\cdots
g^{\alpha_{n-1}\alpha_n)}\mu_{\alpha_n}=g^{(\alpha_1\alpha_2}\cdots
g^{\alpha_{n-1})\alpha_n}\mu_{\alpha_n}=g^{(\alpha_1\alpha_2}\cdots
g^{\alpha_{n-3}\alpha_{n-2}}\mu^{\alpha_{n-1})}.
\end{equation}
When r=2, by using eq. $(\ref{15r-4.3})$ we have
\begin{eqnarray*}
&&g^{(\alpha_1\alpha_2}\cdots
g^{\alpha_{n-1}\alpha_n)}\mu_{\alpha_{n-1}}\mu_{\alpha_n}\\
&=&\frac{1}{n-1}\left[g^{(\alpha_1\alpha_2}\cdots
g^{\alpha_{n-3}\alpha_{n-2})}\mu^{\alpha_{n-1}}+ (n-2)g^{\alpha_{n-1}(\alpha_1}\cdots
g^{\alpha_{n-4}\alpha_{n-3}}\mu^{\alpha_{n-2})}\right]\mu_{\alpha_{n-1}}=\\
&=& \frac{-\gamma^2}{n-1}g^{(\alpha_1\alpha_2}\cdots g^{\alpha_{n-3}\alpha_{n-2})}+
\frac{n-2}{n-1}g^{\alpha_{1}\alpha_2}\cdots
g^{\alpha_{n-5}\alpha_{n-4}}\mu^{\alpha_{n-3}}\mu_{\alpha_{n-2})}
\end{eqnarray*}
from which $(\ref{15r-4.1})$ and $(\ref{15r-4.2})$ with r=2.\\
Let us assume now that our proposition holds up to a fixed integer r and let us prove
that it holds also with r+1 instead of r. By multiplying eq. $(\ref{15r-4.1})$ times
$\mu_{\alpha_{n-r}}$ we find
\begin{eqnarray*}
g^{(\alpha_1\alpha_2}\cdots g^{\alpha_{n-1}\alpha_n)}\mu_{\alpha_{n-r}}\cdots
\mu_{\alpha_n}&=&\sum_{S=0}^{\left[\frac{n-r}{2}\right]-1}\phi_{S+1,r}^{n-r}\frac{2S+2}{n-r}g^{(\alpha_1\alpha_2}\cdots
g^{\alpha_{2S-1}\alpha_{2S}}\mu^{\alpha_{2S+1}}\cdots \mu^{\alpha_{n-r-1})}+\\
&+&\sum_{s=0}^{\left[\frac{n-r}{2}\right]}\phi_{s,r}^{n-r}\frac{n-r-2s}{n-r}g^{(\alpha_1\alpha_2}\cdots
g^{\alpha_{2s-1}\alpha_{2s}}\mu^{\alpha_{2s+1}}\cdots \mu^{\alpha_{n-r-1})}(-\gamma^2),
\end{eqnarray*}
where, in the first summation, we have put s=S+1.\\
It follows $(\ref{15r-4.1})$ with r+1 instead of r and
\begin{equation*}
\phi^{n-r-1}_{s,r+1}=
  \begin{cases}
    \phi^{n-r}_{s+1,r}\frac{2s+2}{n-r}-\gamma^2\phi^{n-r}_{s,r}\frac{n-r-2s}{n-r} & \text{for $s=0,\cdots,\left[\frac{n-r}{2}\right]-1$}, \\
     & \text{and for $s=\left[\frac{n-r}{2}\right]$ in the case r even},\\
     ~\\
    -\gamma^2\phi^{n-r}_{s,r}\frac{n-r-2s}{n-r} & \text{for $s=\frac{n-r-1}{2}$ in the case r odd}.
  \end{cases}
\end{equation*}
We have taken into account that n-r-2s=0 when $s=\left[\frac{n-r}{2}\right]$ and r is
even.\\
By using this result and eq. $(\ref{15r-4.2})$ we have that
\begin{itemize}
  \item For $0\leq s\leq \frac{n}{2}-r-2$, both $\phi^{n-r}_{s+1,r}$ and $\phi^{n}_{s,r}$
  are zero, from which $\phi^{n-r-1}_{s,r+1}=0$.
  \item For $s=\frac{n}{2}-r-1$, we have $\phi^{n-r}_{s,r}=0$ and
  $\phi^{n-r-1}_{s,r+1}=\frac{(n-r-1)!}{(n-2r-2)!!(n-1)!!}$.
  \item For $s=\frac{n}{2}-r,\cdots,\left[\frac{n-r}{2}\right]-1$ and for $s=\left[\frac{n-r}{2}\right]$ in the case r even  we have
  \begin{equation*}
  \phi^{n-r-1}_{s,r+1}=\frac{(r+1)!~(n-r-1)!}{(2s+2r+2-n)!!~(2s)!!~(n-r-1-2s)!~(n-1)!!}~(-\gamma^2)^{s+r+1-\frac{n}{2}}.
  \end{equation*}
  \item For $s=\frac{n-r-1}{2}$ and r odd  we have
  $\phi^{n-r-1}_{s,r+1}=\frac{(r)!~(n-r-1)!}{(r-1)!!~(n-r-1)!!~(n-1)!!}~(-\gamma^2)^{\frac{r+1}{2}}$.
\end{itemize}
In this way we have obtained $(\ref{15r-4.2})$ with r+1 instead of r as we desired to
prove.
\end{proof}
\section{The case N=1.}
When N=1 only equation $(\ref{8})$ with k=0 has to be exploited.\\
We prove now that it amounts to giving the following expression
for the leading term of $C^{\alpha A_1\cdots A_h}$
\begin{eqnarray}\label{asterisco}
&&C^h_{\frac{M}{2}h}=\gamma^{-6}\sum_{q=0}^{\frac{Mh-2}{2}}(-m^2)^{\frac{M}{2}h}\frac{d^h
c_{h,q}(\lambda)}{d\lambda^h}\left(\frac{1}{\gamma^2}\right)^q\frac{(Mh+1)!!}{(Mh-2q-2)!!}\nonumber\\
~\nonumber\\
\text{with}\quad && c_{h,q}=c_{h-1,q} \quad \text{for $q=0,\cdots,\frac{M(h-1)-2}{2}$}\nonumber\\
\text{and}\quad && c_{h,q}\quad \text{for $q=\frac{M(q-1)}{2},\cdots, \frac{Mh-2}{2}$ are
arbitrary functions of $\lambda$}.
\end{eqnarray}
Let us prove this with the iterative procedure.\\
It holds for h=0, because in its right hand side there are no terms, and its left hand
side is zero for eqs. $(\ref{8d})$ and $(\ref{4})$.\\
Let us apply now the Proposition \ref{prop7} to eq. $(\ref{8})$ with k=0,
$r=\frac{M}{2}$, m=Mh+1. We find that
\begin{eqnarray*}
&&C^{h+1}_{\frac{M}{2}(h+1)}=\gamma^{-6}\sum_{q=0}^{\frac{Mh-2}{2}}(-m^2)^{\frac{M}{2}(h+1)}\frac{d^{h+1}
c_{h,q}(\lambda)}{d\lambda^{h+1}}\left(\frac{1}{\gamma^2}\right)^q\frac{(Mh+1)!!}{(Mh-2q-2)!!}\frac{[M(h+1)+1]!}{(Mh+1)!}
\\ && \frac{(Mh)!!}{[M(h+1)]!!}\frac{(Mh-2q-2)!!}{[M(h+1)-2q-2]!!}+
\sum_{i=0}^{\frac{M-2}{2}}f_{i,\frac{M}{2}}\gamma^{-2(3+\frac{hM}{2}+i)}
\end{eqnarray*}
which is eq. $(\ref{asterisco})_1$ with h+1 instead of h ($c_{h+1,q}=c_{h,q}$ has to be
used), $i=q-\frac{Mh}{2}$,
\begin{equation*}
c_{h+1,q}=
  \begin{cases}
    c_{h,q} & \text{for $q=0,\cdots,\left[\frac{Mh-2}{2}\right]$}, \\
    \frac{[M(h+1)-2q-2]!!}{[M(h+1)+1]!!}(-m^2)^{-\frac{M}{2}(h+1)}f^{*}_{q-\frac{Mh}{2},\frac{M}{2}} & \text{for $q=\frac{Mh}{2},\cdots,\frac{M(h+1)-2}{2}$ }.
  \end{cases}
\end{equation*}
with $f^{*}_{q-\frac{Mh}{2},\frac{M}{2}}$ defined by
$\frac{d^{h+1}}{d \lambda^{h+1}
}f^{*}_{q-\frac{Mh}{2},\frac{M}{2}}=f_{q-\frac{Mh}{2},\frac{M}{2}}$.\\
We note that from eq. $(\ref{asterisco})_2$ it follows
\begin{equation}\label{9.1}
c_{h,q}=c_{h-j,q}\qquad \qquad \text{for $q=0,\cdots,\frac{M(h-j)-2}{2}$}.
\end{equation}
Also this relation can be proved with the iterative procedure. It holds when j=0. From
$(\ref{9.1})$ and $(\ref{asterisco})_2$ it follows $c_{h,q}=c_{h-j,q}=c_{h-j-1,q}$ for
$q=0,\cdots,\frac{M(h-j-1)-2}{2}$ that is eq. $(\ref{9.1})$ with j+1 instead of j, as we
desired to prove.\\
Let us search now the other coefficients $C_s^h$. To this end, it
is better to prove firstly that from eq. $(\ref{triangolo})$ with
n=Mh+1 it follows
\begin{equation}\label{9r-2.1}
\phi_{s-r}^{Mh+1}=(-4)^r\frac{s!}{(s-r)!}\frac{(Mh+1-2s)!}{(Mh+1-2s+2r)!}\frac{\partial^r}{\partial(\gamma^2)^r}\phi_s^{Mh+1}.
\end{equation}
Let us prove this with the iterative procedure.\\
It holds for r=0. Let us assume that it holds up to the index r. From eq.
$(\ref{triangolo})$ with n=Mh+1 and s-r instead of s, we find
\begin{eqnarray*}
\phi_{s-r-1}^{Mh+1}&=&\frac{-2s+2r}{\gamma}\frac{1}{(Mh+3-2s+2r)(Mh+2-2s+2r)}\frac{\partial}{\partial\gamma}\phi_{s-r}^{Mh+1}=\\
&=&-4 (s-r)\frac{1}{(Mh+3-2s+2r)(Mh+2-2s+2r)}\frac{\partial}{\partial(\gamma^2)}\phi_{s-r}^{Mh+1}=\\
&=&(-4)^{r+1}\frac{s!}{(s-r-1)!}
\frac{(Mh+1-2s)!}{(Mh+3-2s+2r)!}\frac{\partial^{r+1}}{\partial(\gamma^2)^{r+1}}\phi_{s}^{Mh+1},
\end{eqnarray*}
where in the last passage eq. $(\ref{9r-2.1})$ has been used. In
this way we have proved that $(\ref{9r-2.1})$ holds when r+1
replaces r; this completes the proof of eq.
$(\ref{9r-2.1})$ and we use now it to find $C_s^h$.\\
If we write eq. $(\ref{9r-2.1})$ with $s=\frac{M}{2}h$ and $r=s-s^{*}$, jointly to eq.
$(\ref{asterisco})$, we find that
\begin{eqnarray*}
C^h_{s^{*}}&=&
(-4)^{\frac{M}{2}h-s^{*}}\frac{\left(\frac{M}{2}h\right)!}{(s^{*})!}
\frac{1}{(Mh+1-2s^{*})!}\sum_{q=0}^{\frac{Mh-2}{2}}(-m^2)^{\frac{M}{2}h}\cdot\\
&\cdot & \frac{d^h
c_{h,q}(\lambda)}{d\lambda^h}\frac{(Mh+1)!!}{(Mh-2-2q)!!}(-1)^{\frac{M}{2}h-s^{*}}\frac{(q+2+\frac{M}{2}h-s^{*})!}{(q+2)!}
\left(\frac{1}{\gamma^2}\right)^{q+3+\frac{M}{2}h-s^{*}},
\end{eqnarray*}
that is, the above mentioned eq. $(\ref{r5-1.4})$.
\section{The case $N>1$}
In this case we have to impose eqs. $(\ref{8})$ and $(\ref{9})$. But it will be firstly
useful to transform them into more easy equations. To this end we may use the notation
$\widetilde{B}_i=\alpha_{i_1}\cdots\alpha_{i_{N-1}}$ (similar to the already used
multindex notation $B_i=\alpha_{i_1}\cdots\alpha_{i_{N}}$) and Proposition \ref{prop4};
we obtain that the tensor $D_{h,k}^{\alpha A_1\cdots A_h \widetilde{B}_1\cdots
\widetilde{B}_k}$ exists, such that
\begin{equation}\label{stella}
C_{h,k}^{\alpha A_1\cdots A_h
\beta_1\widetilde{B}_1\cdots\beta_k\widetilde{B}_k}=\frac{\partial^k}{\partial\mu_{\beta_1}
\cdots\partial\mu_{\beta_k}} D_{h,k}^{\alpha A_1\cdots A_h
\widetilde{B}_1\cdots \widetilde{B}_k}.
\end{equation}
After that eq. $(\ref{9})$ becomes
\begin{equation*}
\frac{\partial^{k+1}}{\partial\mu_{\beta_1}\cdots\partial\mu_{\beta_k}\partial\mu_{\beta}}D_{h,k+1}^{\alpha
A_1\cdots A_h \widetilde{B}_1\cdots
\widetilde{B}_k\gamma_1\cdots\gamma_{N-1}}g_{\gamma_1\gamma_2}\cdots
g_{\gamma_{N-2}\gamma_{N-1}}=\frac{\partial^{k+1}}{\partial\mu_{\beta}\partial\mu_{\beta_1}\cdots\partial\mu_{\beta_k}}
D_{h,k}^{\alpha A_1\cdots A_h \widetilde{B}_1\cdots
\widetilde{B}_k}(-m^2)^{\frac{N-1}{2}}.
\end{equation*}
For the Proposition \ref{prop4} with n=hM+k(N-1)+1 the general solution of this equation
is
\begin{eqnarray}\label{oo}
 D_{h,k+1}^{\alpha A_1\cdots A_h \widetilde{B}_1\cdots
\widetilde{B}_k\gamma_1\cdots\gamma_{N-1}}g_{\gamma_1\gamma_2}\cdots
g_{\gamma_{N-2}\gamma_{N-1}}=D_{h,k}^{\alpha A_1\cdots A_h
\widetilde{B}_1\cdots
\widetilde{B}_k}(-m^2)^{\frac{N-1}{2}}+\nonumber \\
+\sum_{i=0}^{\left[\frac{k-1}{2}\right]}\phi_{i,h,k}(\lambda)
g^{(\alpha A_1\cdots A_h \widetilde{B}_1\cdots \widetilde{B}_k
\alpha_{hM+k(N-1)+2}\cdots
\alpha_{hM+k(N-1)+2i})}\mu_{\alpha_{hM+k(N-1)+2}}\cdots
\mu_{\alpha_{hM+k(N-1)+2+2i}}.
\end{eqnarray}
Let us transform now the unknown tensors $D_{h,k}^{\cdots}$ into the tensors
$E_{h,k}^{\cdots}$, according to the following rule
\begin{eqnarray}\label{ooo}
D_{h,k}^{\alpha A_1\cdots A_h \widetilde{B}_1\cdots
\widetilde{B}_k}= E_{h,k}^{\alpha
A_1\cdots A_h \widetilde{B}_1\cdots \widetilde{B}_k}+\nonumber \\
+\sum_{i=0}^{\left[\frac{k-2}{2}\right]}\psi_{i,h,k}(\lambda)
g^{(\alpha A_1\cdots A_h \widetilde{B}_1\cdots \widetilde{B}_k
\alpha_{hM+k(N-1)+2}\cdots
\alpha_{hM+k(N-1)+2i})}\mu_{\alpha_{hM+k(N-1)+2}}\cdots
\mu_{\alpha_{hM+k(N-1)+2+2i}}.
\end{eqnarray}
with $\psi_{i,h,k}(\lambda)$ defined with the iterative method by
\begin{eqnarray*}
 \psi_{i,h,0}&=&0, \\
 \psi_{i,h,1}&=&0, \\
 \psi_{i,h,k+1}&=&\frac{[Mh+k(N-1)+2i+4]!!}{[hM+(k+1)(N-1)+2i+4]!!}\frac{[hM+(k+1)(N-1)+2i+1]!!}{[hM+k(N-1)+2i+1]!!}\\
 && \left(\psi_{i,h,k}(-m^2)^{\frac{N-1}{2}}+\phi_{i,h,k}\right) \\
 && \text{in the case k even and $i=0,\cdots,\frac{k-2}{2}$ and in the case k odd and
 $i=0,\cdots,\frac{k-3}{2}$}\\
 ~\\
\psi_{i,h,k+1}&=&\frac{[Mh+k(N-1)+2i+4]!!}{[hM+(k+1)(N-1)+2i+4]!!}\frac{[hM+(k+1)(N-1)+2i+1]!!}{[hM+k(N-1)+2i+1]!!}
 \phi_{i,h,k} \\
 && \text{in the case k odd and $i=\frac{k-1}{2}$}.
\end{eqnarray*}
We note that also $E_{h,k}^{\cdots}\in\mathcal{F}$.\\
Thanks to this transformation, eq. $(\ref{oo})$ becomes
\begin{eqnarray}\label{b.1}
E_{h,k+1}^{\alpha A_1\cdots A_h \widetilde{B}_1\cdots
\widetilde{B}_k\gamma_1\cdots\gamma_{N-1}}g_{\gamma_1\gamma_2}\cdots
g_{\gamma_{N-2}\gamma_{N-1}}=E_{h,k}^{\alpha A_1\cdots A_h \widetilde{B}_1\cdots
\widetilde{B}_k}(-m^2)^{\frac{N-1}{2}}.
\end{eqnarray}
Let us note how many 4-vectors $\mu_{\alpha}$ intervene in the
second term in the right hand side of $(\ref{ooo})$; they are
$2i+1\leq 2\left[\frac{k-2}{2}\right]+1\leq k-2+1<k$; therefore
eq. $(\ref{oo})$, substituted in $(\ref{stella})$ transforms it
into
\begin{equation}\label{b.2}
C_{h,k}^{\alpha A_1\cdots A_h \beta_1\widetilde{B}_1\cdots
\beta_k\widetilde{B}_k}=\frac{\partial^k}{\partial\mu_{\beta_1}\cdots\partial\mu_{\beta_k}}E_{h,k}^{\alpha
A_1\cdots A_h \widetilde{B}_1\cdots \widetilde{B}_k}.
\end{equation}
Eqs. $(\ref{b.1})$ and $(\ref{b.2})$ substitute now eqs. $(\ref{stella})$ and
$(\ref{oo})$; consequently, these last one can now be left out!\\
Eq. $(\ref{b.2})$ with k=0 now yields
\begin{equation}\label{c.2}
C_{h,0}^{\alpha A_1\cdots A_h}=E_{h,0}^{\alpha A_1\cdots A_h}
\end{equation}
which is restricted by eq. $(\ref{8})$ with k=0. After that eq. $(\ref{b.1})$ will give,
with an iterative procedure, $E_{h,k+1}^{\cdots}$. Let us firstly deduce, from this
procedure, some properties.
\begin{property}
The leading term of $E_{h,0}^{\alpha A_1\cdots A_h}$ is $\left(\frac{1}{\gamma}\right)^6$
multiplied by a polynomial in the variable $\frac{1}{\gamma^2}$.
\end{property}
\begin{proof}
This property is evident from eq. $(\ref{asterisco})$.
\end{proof}
\begin{property}
The leading term of $E_{h,k}^{\alpha A_1\cdots A_h \widetilde{B}_1\cdots
\widetilde{B}_k}$ is $\left(\frac{1}{\gamma}\right)^6$ multiplied by a polynomial in the
variable $\frac{1}{\gamma^2}$.
\end{property}
\begin{proof}
Let us prove this with the iterative procedure.\\
It holds when k=0, for property 1.\\
Let us assume it up to the index k. From eq. $(\ref{b.1})$, by applying Proposition
\ref{prop7} with m=Mh+(N-1)k+1, $r=\frac{N-1}{2}$, we find that property 2 holds also
when k+1 replaces k. \\
This completes its proof.
\end{proof}\\
By substituting now eq. $(\ref{b.2})$ into eq. $(\ref{8})$, this last one becomes
\begin{equation*}
\frac{\partial^k}{\partial\mu_{\beta_1}\cdots\partial\mu_{\beta_k}}\left[
E_{h+1,k}^{\alpha A_1\cdots A_h \widetilde{B}_1\cdots\widetilde{B}_k
\gamma_1\gamma_2\cdots\gamma_{M-1}\gamma_M}g_{\gamma_1\gamma_2}\cdots
g_{\gamma_{M-1}\gamma_M}-(-m^2)^{\frac{M}{2}}\frac{\partial}{\partial\lambda}
E_{h,k}^{\alpha A_1\cdots A_h \widetilde{B}_1\cdots\widetilde{B}_k}\right]=0
\end{equation*}
For Proposition \ref{prop4} with n=Mh+(N-1)k+1, r=k, it follows that
\begin{eqnarray}\label{c.1}
&&E_{h+1,k}^{\alpha A_1\cdots A_h \widetilde{B}_1\cdots\widetilde{B}_k
\gamma_1\gamma_2\cdots\gamma_{M-1}\gamma_M}g_{\gamma_1\gamma_2}\cdots
g_{\gamma_{M-1}\gamma_M}-(-m^2)^{\frac{M}{2}}\frac{\partial}{\partial\lambda}
E_{h,k}^{\alpha A_1\cdots A_h \widetilde{B}_1\cdots\widetilde{B}_k}=\nonumber\\
&&= \sum_{i=0}^{\left[\frac{k-2}{2}\right]}\phi_i(\lambda)g^{(\alpha A_1\cdots A_h
\widetilde{B}_1\cdots\widetilde{B}_k\alpha_1\cdots\alpha_{2i+1})}\mu_{\alpha_1}\cdots\mu_{\alpha_{2i+1}}.
\end{eqnarray}
where the notation $g^{\alpha_1\cdots\alpha_{2n}}=g^{(\alpha_1\alpha_2}\cdots
g^{\alpha_{2n-1}\alpha_{2n})}$ has been used.\\
But, for the above property 2, jointly with the Propositions \ref{prop6} and \ref{prop8},
we have that the left hand side of eq. $(\ref{c.1})$ has a leading term of the type
$\left(\frac{1}{\gamma}\right)^6$ multiplied by a polynomial in the variable
$\frac{1}{\gamma^2}$, while its right hand side has a leading term polynomial in
$\gamma^2$. It follows necessarily that both these leading terms are zero; in other
words, both sides of eq. $(\ref{c.1})$ are zero. In particular, from the left hand side
we obtain
\begin{equation}\label{d.1}
E_{h+1,k}^{\alpha A_1\cdots A_h \widetilde{B}_1\cdots\widetilde{B}_k
\gamma_1\gamma_2\cdots\gamma_{M-1}\gamma_M}g_{\gamma_1\gamma_2}\cdots
g_{\gamma_{M-1}\gamma_M}=(-m^2)^{\frac{M}{2}}\frac{\partial}{\partial\lambda}
E_{h,k}^{\alpha A_1\cdots A_h \widetilde{B}_1\cdots\widetilde{B}_k}.
\end{equation}
Therefore, we have to impose only eqs. $(\ref{b.1})$ and
$(\ref{d.1})$; after that, the tensor $C_{h,k}^{\alpha A_1\cdots
A_h B_1\cdots B_k}$ defined by eq. $(\ref{b.2})$ is the more
general solution of eq. $(\ref{8})$ and $(\ref{9})$.
\subsection{A consequence of eqs. $(\ref{b.1})$ and $(\ref{d.1})$}
We will see now that, as a consequence of eqs. $(\ref{b.1})$ and $(\ref{d.1})$, we may
find the leading term of $E_{h,k}^{\alpha A_1\cdots A_h
\widetilde{B}_1\cdots\widetilde{B}_k}$ except for a set of arbitrary functions of the
single variable $\lambda$. This result is the subsequent $(\ref{e2.1})$.\\
To this end let us firstly prove that
\begin{equation}\label{d.2}
E^{0,k}_{\frac{N-1}{2}k}=\gamma^{-6}\sum_{q=0}^{\frac{(N-1)k-2}{2}}(-m^2)^{\frac{N-1}{2}k}
c_{k,q}(\lambda)\frac{[(N-1)k+1]!!}{[(N-1)k-2q]!!}[(N-1)k-2q]\left(\frac{1}{\gamma^2}\right)^q,
\end{equation}
where the last N-1 functions $c_{k,q}$ are arbitrary functions of $\lambda$ and the
remainder are
\begin{equation}\label{d.3}
c_{k,q}=c_{k-1,q}.
\end{equation}
Let us prove it with the iterative procedure.\\
It holds when k=0, because in this case the right hand side has no terms, while the left
hand side is zero for eqs. $(\ref{c.2})$ with h=0, $(\ref{8d})$ and $(\ref{4})$.\\
Let us apply Proposition \ref{prop7} to eq. $(\ref{b.1})$ with h=0, $r=\frac{N-1}{2}$,
m=k(N-1)+1. We find that
\begin{eqnarray*}
&&E^{0,k+1}_{\frac{N-1}{2}(k+1)}=\gamma^{-6}\sum_{q=0}^{\frac{(N-1)k-2}{2}} c_{k,q}
\left(\frac{1}{\gamma^2}\right)^q \frac{[(N-1)k+1]!!}{[(N-1)k-2q]!!}[(N-1)k-2q]\cdot\\
&&\cdot(-m^2)^{\frac{N-1}{2}(k+1)} \frac{[(N-1)k+1+N-1]!}{[(N-1)k+1]!}
\frac{[(N-1)k]!!}{[(N-1)k+N-1]!!} \frac{[(N-1)k-2q-2]!!}{[(N-1)(k+1)-2q-2]!!}+\\
&&+\sum_{i=0}^{\frac{N-3}{2}}f_{i,\frac{N-1}{2}}(\lambda)\gamma^{-[6+k(N-1)+2i]}.
\end{eqnarray*}
Then we have found eq. $(\ref{d.2})$ with k+1 instead of k, $i=q-k\frac{N-1}{2}$ and
\begin{equation*}
c_{k+1,q}=
  \begin{cases}
    c_{k,q} & \text{for $q=0,\cdots,\frac{(N-1)k-2}{2}$}, \\
    f_{q-k\frac{N-1}{2},\frac{N-1}{2}}\frac{[(N-1)(k+1)-2q-2]!!}{[(N-1)(k+1)+1]!}(-m^2)^{-\frac{N-1}{2}(k+1)} & \text{for $q=\frac{(N-1)k}{2},\cdots,\frac{(N-1)(k+1)-2}{2}$}.
  \end{cases}
\end{equation*}
This completes the proof.\\
It will be useful in the sequel also the
\begin{property}
$c_{k,q}=c_{k-i,q}$ for $q=0,\cdots \frac{(N-1)(k-i)-2}{2}$.
\end{property}
\begin{proof}
Let us prove this with the iterative procedure.\\
It is obvious when i=0.\\
From eq. $(\ref{d.3})$ with k-i instead of k, we have
\begin{equation*}
c_{k,q}=c_{k-i,q}=c_{k-i-1,q}
\end{equation*}
for $q=0,\cdots \frac{(N-1)(k-i-1)-2}{2}$. So the property is valid also when i+1
replaces i.
\end{proof}\\
Now, from eq. $(\ref{b.1})$ it follows
\begin{equation*}
E_{h,k}^{\alpha A_1\cdots A_h \widetilde{B}_1\cdots \widetilde{B}_k}= E_{h,k+2}^{\alpha
A_1\cdots A_h \widetilde{B}_1\cdots \widetilde{B}_k
\widetilde{B}_{k+1}\widetilde{B}_{k+2}}
g_{\widetilde{B}_{k+1}}g_{\widetilde{B}_{k+2}}(-m^2)^{-(N-1)},
\end{equation*}
and so on, until
\begin{equation*}
E_{h,k}^{\alpha A_1\cdots A_h \widetilde{B}_1\cdots \widetilde{B}_k}= E_{h,k+Mh}^{\alpha
A_1\cdots A_h \widetilde{B}_1\cdots \widetilde{B}_{k}\widetilde{B}_{k+1}\cdots
\widetilde{B}_{k+Mh}} g_{\widetilde{B}_{k+1}}\cdots
g_{\widetilde{B}_{k+Mh}}(-m^2)^{-\frac{(N-1)Mh}{2}}.
\end{equation*}
By applying h times eq. $(\ref{d.1})$, the above equation becomes
\begin{equation}\label{3.}
E_{h,k}^{\alpha A_1\cdots A_h \widetilde{B}_1\cdots \widetilde{B}_k}= \frac{\partial^h}
{\partial\lambda^h}E_{0,k+Mh}^{\alpha A_1\cdots A_h \widetilde{B}_1\cdots
\widetilde{B}_{k}\gamma_1\cdots \gamma_{Mh(N-2)}} g_{\gamma_1\gamma_2}\cdots
g_{\gamma_{Mh(N-2)-1}\gamma_{Mh(N-2)}}(-m^2)^{-\frac{(N-2)Mh}{2}}.
\end{equation}
In this way all the $E_{h,k}^{\cdots}$ are determined in terms of $E_{0,k}^{\cdots}$.\\
From eqs. $(\ref{3.})$, $(\ref{d.2})$ and Proposition \ref{prop6} with
n=Mh+(N-1)k+1+Mh(N-2) and $r=\frac{M}{2}h(N-2)$ we find that
\begin{eqnarray*}
&&
E^{h,k}_{\frac{Mh+(N-1)k}{2}}=\gamma^{-6}\sum_{q=0}^{\frac{(N-1)(k+Mh)-2}{2}}(-m^2)^{\frac{N-1}{2}k+\frac{Mh}{2}}
  \frac{d^h c_{k+Mh,q}}{d\lambda^h}\cdot
\left(\frac{1}{\gamma^2}\right)^q \frac{[(N-1)(k+Mh)+1]!!}{[(N-1)(k+Mh)-2q-2]!!}\\
&&
\frac{[Mh+(N-1)k+1]!!}{[(Mh+k)(N-1)+1]!!}\eta[k(N-1)+Mh-2q,(Mh+k)(N-1)-2-2q].
\end{eqnarray*}
This expression can be simplified because, for $q\geq\frac{Mh+k(N-1)}{2}$ we have
$Mh+(N-1)k-2q\leq 0$, $(Mh+k)(N-1)-2-2q\geq 0$, so that at least one, among the factors
intervening in $\eta(\cdots,\cdots)$, is zero. Then we can restrict to the values with
$q\leq\frac{Mh+k(N-1)-2}{2}$ and, consequently,
\begin{equation}\label{e2.1}
E^{h,k}_{\frac{Mh+(N-1)k}{2}}=\gamma^{-6}\sum_{q=0}^{\frac{(N-1)k+Mh-2}{2}}(-m^2)^{\frac{N-1}{2}k+\frac{Mh}{2}}
\frac{d^h c_{k+Mh,q}}{d\lambda^h} \left(\frac{1}{\gamma^2}\right)^q
\frac{[Mh+(N-1)k+1]!!}{[(N-1)k+Mh-2q-2]!!}.
\end{equation}
This agrees with the above used property 2.
\subsection{Equivalence of eqs. $(\ref{b.1})$ and $(\ref{d.1})$ with eq. $(\ref{e2.1})$}
The result $(\ref{e2.1})$ has been proved as a consequence of eqs.
$(\ref{b.1})$ and $(\ref{d.1})$. We prove now the vice versa,
i.e., that eqs. $(\ref{b.1})$ and
$(\ref{d.1})$ become identities when eq. $(\ref{e2.1})$ is used. \\
Let us begin with eq. $(\ref{b.1})$: For the Proposition \ref{prop6} with
n=hM+(k+1)(N-1)+1 and $r=\frac{N-1}{2}$, and $(\ref{e2.1})$, the leading term of the left
hand side of eq. $(\ref{b.1})$ is
\begin{eqnarray*}
\gamma^{-6}\sum_{q=0}^{\frac{(N-1)(k+1)+Mh-2}{2}}
\frac{[(N-1)(k+1)+Mh+1]!!}{[(N-1)(k+1)+Mh-2q-2]!!}(-m^2)^{\frac{N-1}{2}(k+1)+\frac{Mh}{2}}
\frac{d^h c_{k+1+Mh,q}}{d\lambda^h}\cdot \\
\left(\frac{1}{\gamma^2}\right)^q \frac{[(N-1)k+Mh+1]!!}{[(N-1)(k+1)+Mh+1]!!}
\eta(Mh+k(N-1)-2q,Mh+(k+1)(N-1)-2-2q).
\end{eqnarray*}
If $q\geq \frac{Mh+k(N-1)}{2}$ we have $hM+k(N-1)-2q\leq 0$, $hM+(k+1)(N-1)-2-2q\geq 0$,
so that $\eta(\cdots,\cdots)=0$; therefore we can limit to values with $a\leq
\frac{Mh+k(N-1)}{2}-1$ and the above leading term becomes equal to the right hand side of
eq. $(\ref{e2.1})$ premultiplied by $(-m^2)\frac{N-1}{2}$ (use the property $(\ref{d.3})$
with k+1+Mh instead of k, i.e., $c_{k+1+Mh,q}=c_{k+Mh,q}$ for
$q=0,\cdots,\frac{Mh+k(N-1)-2}{2}\leq (N-1)(k+1+Mh)-2$).\\
Therefore, eq. $(\ref{b.1})$ is an identity.\\
Let us now prove the same thing for $(\ref{d.1})$: For the Proposition \ref{prop6} with
n=M(h+1)+k(N-1)+1 and $r=\frac{M}{2}$, and $(\ref{e2.1})$, the leading term of the left
hand side of eq. $(\ref{d.1})$ is
\begin{eqnarray*}
\gamma^{-6}\sum_{q=0}^{\frac{(N-1)k+M(h+1)-2}{2}}\frac{[(N-1)k+M(h+1)+1]!!}{[(N-1)k+M(h+1)-2q-2]!!}
(-m^2)^{\frac{N-1}{2}k+\frac{M(h+1)}{2}}
\frac{d^{h+1} c_{k+Mh+M,q}}{d\lambda^{h+1}}\cdot \\
\left(\frac{1}{\gamma^2}\right)^q \frac{[(N-1)k+Mh+1]!!}{[(N-1)k+M(h+1)+1]!!}
\eta(Mh+k(N-1)-2q,M(h+1)+k(N-1)-2-2q).
\end{eqnarray*}
When $q\geq \frac{Mh+k(N-1)}{2}$ we have $\eta(\cdots,\cdots)=0$
so that there remain terms with $q\leq \frac{Mh+k(N-1)-2}{2}$,
which is the right hand side of eq. $(\ref{e2.1})$ multiplied by
$(-m^2)^{\frac{M}{2}}$ and derived with respect to $\lambda$ (use
$c_{k+Mh+M,q}=c_{k+Mh,q}$ for $q=0,\cdots,\frac{Mh+k(N-1)-2}{2}$
which holds for property 3 written with i=M and k+Mh+M instead of
k because $Mh+k(N-1)-2\leq (N-1)(k+Mh)-2$). Therefore, eq.
$(\ref{d.1})$ is an identity.
\subsection{Determination of the tensor $C_{h,k}^{\alpha A_1\cdots A_h B_1\cdots B_k}$}
After having imposed eqs. $(\ref{b.1})$ and $(\ref{d.1})$, let us
now impose eq. $(\ref{b.2})$ for the determination of the tensor
$C_{h,k}^{\alpha A_1\cdots A_h B_1\cdots B_k}$. To this end, let
us firstly note that from eq. $(\ref{triangolo2})_2$ with n odd
and $s=\frac{n+1}{2}$ we find that the leading term of
$\frac{\partial\phi^{\alpha_1\cdots\alpha_n}}{\partial\mu_{\beta}}$
is
\begin{equation}\label{e.4-1}
\phi^{n+1}_{\frac{n+1}{2}}=\phi^n_{\frac{n-1}{2}}.
\end{equation}
From eq. $(\ref{triangolo2})_2$ with n+1 instead of n and $s=\frac{n+1}{2}$ we find that
the leading term of $\frac{\partial^2
\phi^{\alpha_1\cdots\alpha_n}}{\partial\mu_{\beta_1}\partial\mu_{\beta_2}}$ is
\begin{equation*}
\phi^{n+2}_{\frac{n+1}{2}}=2\frac{n+2}{n+1}\phi^{n+1}_{\frac{n-1}{2}}=
-\frac{n+2}{\gamma}\frac{\partial}{\partial\gamma}\phi^n_{\frac{n-1}{2}}
\end{equation*}
where in the last passage eq. $(\ref{e.4-1})$ and eq.
$(\ref{triangolo})$ with n+1
instead of n and $s=\frac{n+1}{2}$ have been used.\\
If $\phi^{n}_{\frac{n-1}{2}}$ depends on $\gamma$ by means of $\gamma^2$, it follows that
$\phi^{n+2}_{\frac{n+1}{2}}$ also depends on $\gamma$ by means of $\gamma^2$ and is
\begin{equation}\label{e.4-2}
\phi^{n+2}_{\frac{n+1}{2}}=-2(n+2)\frac{\partial}{\partial\gamma^2}\phi^n_{\frac{n-1}{2}}
\end{equation}
It follows that
\begin{equation}\label{e.4-3}
\phi^{n+2r}_{\frac{n+2r-1}{2}}=(-2)^r\frac{(n+2r)!!}{n!}\frac{\partial^r}{\partial(\gamma^2)^r}\phi^n_{\frac{n-1}{2}}.
\end{equation}
(In fact, it holds for r=0. Let us assume that it holds up to an integer r. Eq.
$(\ref{e.4-2})$ with n+2r instead of n becomes
\begin{equation*}
\phi^{n+2r+2}_{\frac{n+2r+1}{2}}=-2(n+2r+2)\frac{\partial}{\partial\gamma^2}\phi^{n+2r}_{\frac{n+2r-1}{2}}=
(-2)^{r+1}\frac{(n+2r+2)!!}{n!}\frac{\partial^{r+1}}{\partial(\gamma^2)^{r+1}}\phi^{n}_{\frac{n-1}{2}},
\end{equation*}
where $(\ref{e.4-3})$ has been used. The result is again $(\ref{e.4-3})$ but with r+1
instead of r; this completes the proof).\\
From eq. $(\ref{e.4-1})$, with n+2r instead of n, and for eq. $(\ref{e.4-2})$, it follows
\begin{equation*}
\phi^{n+2r+1}_{\frac{n+2r+1}{2}}=
(-2)^{r}\frac{(n+2r)!!}{n!!}\frac{\partial^{r}}{\partial(\gamma^2)^{r}}\phi^{n}_{\frac{n-1}{2}}.
\end{equation*}
This result and eq. $(\ref{e.4-2})$ give
\begin{equation*}
\phi^{n+k}_{\left[\frac{n+k}{2}\right]}=
(-2)^{\left[\frac{k}{2}\right]}\frac{(n+2\left[\frac{k}{2}\right])!!}{n!!}
\frac{\partial^{\left[\frac{k}{2}\right]}}{\partial(\gamma^2)^{\left[\frac{k}{2}\right]}}\phi^{n}_{\frac{n-1}{2}}.
\end{equation*}
This relation, with n=Mh+k(N-1)+1, and eq. $(\ref{e2.1})$ allows
to obtain from eq. $(\ref{b.2})$ that the leading term of
$C_{h,k}^{\alpha A_1\cdots A_h B_1\cdots B_k}$ is
\begin{eqnarray}\label{e.5-1}
C^{h,k}_{\left[\frac{Mh+kN+1}{2}\right]}&=&2^{\left[\frac{k}{2}\right]}\gamma^{-6-2\left[\frac{k}{2}\right]}
\sum_{q=0}^{\frac{Mh+k(N-1)-2}{2}}
\frac{\left(hM+k(N-1)+1+2\left[\frac{k}{2}\right]\right)!!}{[hM+k(N-1)-2q-2]!!}(-m^2)^{\frac{N-1}{2}k+\frac{Mh}{2}}\nonumber\\
&&\frac{d^h c_{k+Mh,q}}{d \lambda^h}
\frac{\left(q+2+\left[\frac{k}{2}\right]\right)!}{(q+2)!}\left(\frac{1}{\gamma^2}\right)^q.
\end{eqnarray}
This result allows to determine the other coefficients $C_s^{h,k}$.\\
To this end, it will be useful to note firstly that, from eq. $(\ref{triangolo})$ with
n=Mh+Nk+1 it follows
\begin{equation}\label{e.5-2}
\phi^n_{s-r}=(-4)^r\frac{s!}{(s-r)!}\frac{(Mh+Nk+1-2s)!}{(Mh+Nk+1-2s+2r)!}
\frac{\partial^r}{\partial(\gamma^2)^r}\phi_s^n.
\end{equation}
(In fact, this relation holds for r=0. Let us assume that it also holds up to an index r;
from eq. $(\ref{triangolo})$  with n=Mh+Nk+1 and s-r instead of s we find that
\begin{eqnarray*}
\phi^n_{s-r-1}&=&\frac{-2s+2r}{\gamma}\frac{1}{(Mh+Nk+3-2s+2r)(Mh+Nk+2-2s+2r)}
\frac{\partial}{\partial\gamma}\phi^n_{s-r}=\\
&=& -4 (s-r)\frac{1}{(Mh+Nk+3-2s+2r)(Mh+Nk+2-2s+2r)}
\frac{\partial}{\partial(\gamma^2)}\phi^n_{s-r}= \\
&=&(-4)^{r+1}\frac{s!}{(s-r-1)!}\frac{(Mh+Nk+1-2s)!}{(Mh+Nk+3-2s+2r)!}
\frac{\partial^{r+1}}{\partial(\gamma^2)^{r+1}}\phi_s^n
\end{eqnarray*}
where in the last passage eq. $(\ref{e.5-2})$ has been used. The result is eq.
$(\ref{e.5-2})$, with r+1 instead of r, and this completes its proof.).\\
Eq. $(\ref{e.5-2})$ with $s=\left[\frac{Mh+Nk+1}{2}\right]$ and
$r=s-s^*$, jointly with eq. $(\ref{e.5-1})$ shows that
\begin{eqnarray*}
&& C_{s^*}^{h,k}=(-4)^{\left[\frac{Mh+Nk+1}{2}\right]-s^*}
\frac{\left[\frac{Mh+Nk+1}{2}\right]!}{s^*!}
\frac{\left(Mh+Nk+1-2\left[\frac{Mh+Nk+1}{2}\right]\right)!}{(Mh+Nk+1-2s^*)!}\\
&& (2)^{\left[\frac{k}{2}\right]}\sum_{q=0}^{\frac{mh+k(N-1)-2}{2}}
\frac{\left(Mh+k(N-1)+1-2\left[\frac{k}{2}\right]\right)!}{(Mh+k(N-1)-2q-2)!}
(-m^2)^{\frac{N-1}{2}k+\frac{Mh}{2}}\\
&& \frac{d^h c_{k+Mh,q}}{d\lambda^h}
\frac{\left(q+2+\left[\frac{k}{2}\right]\right)!}{(q+2)!}
(-1)^{\left[\frac{Mh+Nk+1}{2}\right]-s^*}
\frac{\left(q+2+\left[\frac{k}{2}\right]+\left[\frac{Mh+Nk+1}{2}\right]-s^*\right)!}
{\left(q+2+\left[\frac{k}{2}\right]\right)!}\\
&&\left(\frac{1}{\gamma^2}\right)^{q+3+\left[\frac{k}{2}\right]+\left[\frac{Mh+Nk+1}{2}\right]-s^*}
\end{eqnarray*}
from which the above reported eq. $(\ref{r5-1.2})$, taking into account that
\begin{equation*}
\left(Mh+Nk+1-2\left[\frac{Mh+Nk+1}{2}\right]\right)!=
  \begin{cases}
    0!=1 & \text{if k is odd}, \\
    1!=1 & \text{if k is even},
  \end{cases}
\end{equation*}
and moreover, that
$\left[\frac{Mh+Nk+1}{2}\right]+\left[\frac{k}{2}\right]=\frac{Mh+(N+1)k}{2}$.
\section{Conclusions.}
We consider the present results very satisfactory, because the closure which is found
constitutes a consistent improvement of the previous one appearing in literature. It
takes into account an arbitrary, but fixed, number of moments; moreover it satisfies the
constraints exactly and satisfies also the supplementary conservation law related to the
entropy principle. This fact guarantees the existence and uniqueness, well-posedness and
stability of the solution of the initial value problem. Moreover it allows also further
investigations, such as that concerning the warm plasma ordering and that concerning the
behavior of the subsystems. These may be the subject of future works.

\end{document}